\documentclass{article}
\pdfoutput=1

\usepackage{arxiv}
\usepackage[utf8]{inputenc} 
\usepackage[T1]{fontenc}    
\usepackage{hyperref}       
\hypersetup{
	colorlinks = true,
	linkcolor = black,
	anchorcolor = black,
	citecolor = black,
	filecolor = black,
	urlcolor = blue
}
\usepackage{booktabs}       
\usepackage{amsfonts}       
\usepackage{nicefrac}       
\usepackage{microtype}      
\usepackage{amsmath,bm}
\usepackage{graphicx}
\usepackage[flushleft]{threeparttable}
\usepackage{multirow}
\usepackage{natbib}
\usepackage{amsmath}

\newcommand{\figurehere}[1]{\begin{center}%
	=========================\\%
	Insert Figure #1 about here\\%
	=========================\\%
\end{center}}
\newcommand{\tablehere}[1]{\begin{center}%
	=========================\\%
	Insert Table #1 about here\\%
	=========================\\%
\end{center}}

\title{Obtaining interpretable parameters from reparameterized longitudinal models: transformation matrices between growth factors in two parameter spaces}

\author{
    Jin Liu\thanks{CONTACT Jin Liu. Email: Veronica.Liu0206@gmail.com, \textcircled{c}2021, Journal of Educational and Behavioral Statistics. This paper is not the copy of record and may not exactly replicate the final, authoritative version of the article. Please do not copy or cite without authors' permission. The final article will be available, upon publication, via its DOI: 10.3102/10769986211052009} \\
    Department of Biostatistics\\
    Virginia Commonwealth University \\
    \And
    Robert A. Perera \\
    Department of Biostatistics\\
    Virginia Commonwealth University \\
    \And
    Le Kang \\
    Department of Biostatistics\\
    Virginia Commonwealth University \\
    \And
    Robert M. Kirkpatrick \\
    Department of Psychiatry \\
    Virginia Commonwealth University \\
    \And
    Roy T. Sabo \\
    Department of Biostatistics\\
    Virginia Commonwealth University \\
}

\begin{document}
\maketitle

\begin{abstract}
The linear spline growth model (LSGM), which approximates complex patterns using at least two linear segments, is a popular tool for examining nonlinear change patterns. Among such models, the linear-linear piecewise change pattern is the most straightforward one. An earlier study has proved that other than the intercept and slopes, the knot (or change-point), at which two linear segments join together, can be estimated as a growth factor in a reparameterized longitudinal model in the latent growth curve modeling framework. However, the reparameterized coefficients were no longer directly related to the underlying developmental process and therefore lacked meaningful, substantive interpretation, although they were simple functions of the original parameters. This study proposes transformation matrices between parameters in the original and reparameterized models so that the interpretable coefficients directly related to the underlying change pattern can be derived from reparameterized ones. Additionally, the study extends the existing linear-linear piecewise model to allow for individual measurement occasions, and investigates predictors for the individual-differences in change patterns.  We present the proposed methods with simulation studies and a real-world data analysis. Our simulation studies demonstrate that the proposed method can generally provide an unbiased and consistent estimation of model parameters of interest and confidence intervals with satisfactory coverage probabilities. An empirical example using longitudinal mathematics achievement scores shows that the model can estimate the growth factor coefficients and path coefficients directly related to the underlying developmental process, thereby providing meaningful interpretation. For easier implementation, we also provide the corresponding code for the proposed models.
\end{abstract}


\section{Introduction}\label{Intro}
Longitudinal studies of change are popular in various disciplines to evaluate individual growth over time. If a process under investigation is followed for a long enough time duration, it is likely to exhibit some degree of nonlinear change in which the curve has a nonconstant relationship to recorded time. One possible model to examine nonlinear individual change pattern is a piecewise linear latent growth model \citep{Chou2004PLGC, Harring2006nonlinear, Cudeck2010nonlinear, Kohli2011PLGC, Kohli2013PLGC1, Kohli2013PLGC2, Sterba2014individually}, also known as the linear spline growth model \citep{Grimm2016growth}. By using at least two linear pieces, the linear spline (or piecewise linear) functional form can approximate more complex underlying change patterns. Previous studies have demonstrated a broad application of linear spline trajectory. For example, it has been employed to assess post-surgical rehabilitation \citep{Riddle2015knee, Dumenci2019knee}, the learning process of a specific task \citep{Cudeck2002PLGC}, the onset of alcohol abuse \citep{Li2001PLGC} and intellectual development \citep{Marcoulides2018PLGC}. 

One major challenge of analyzing linear piecewise change patterns is to decide the location of the change-point or `knot' that indicates the occurrence of the change in the developmental process. Although empirical researchers can specify the knot location \textit{a priori} driven by domain knowledge, for example, \citet{Flora2008knot, Sterba2014individually, Riddle2015knee, Dumenci2019knee}, the knot is unknown in most cases. \citet{Marcoulides2018PLGC} proposed to fit a pool of candidate models with a knot at different locations and select the `best' one using the BIC criterion. By building a multi-phase mixed-effects model, \citet{Cudeck2002PLGC} demonstrated that the change-point can also be a random coefficient that allows individuals to transition from one phase to another at different time points. Earlier studies have demonstrated how to estimate knots using frequentist \cite{Cudeck2003knot_F, Harring2006nonlinear, Kwok2010simu, Kohli2011PLGC, Kohli2013PLGC1, Kohli2013PLGC2, Preacher2015repara} and Bayesian \cite{Dominicus2008knot_B, McArdle2008knot_B, Wang2008knot_B, Muniz2011knot_B, Kohli2015PLGC1, Lock2018knot_B} mixed-effects models and growth models.

The most intuitive linear spline function is a bilinear spline growth model (or linear-linear piecewise model). This functional form can identify a process that has two stages with different rates of change \citep{Riddle2015knee, Dumenci2019knee}, but it can also help approximate more complex nonlinear change patterns \citep{Kohli2015PLGC1, Kohli2015PLGC2, Kohli2017PLGC}. \citet{Harring2006nonlinear} developed a linear-linear piecewise model with an unknown fixed knot with a unified functional form of two linear pieces through reparameterization, which is realized by re-expressing growth factors as linear combinations of their original forms. \citet{Kohli2011PLGC}, \citet{Kohli2013PLGC1} and \citet{Grimm2016growth} derived the growth coefficients that are directly related to the underlying change patterns from the estimated coefficients using \textit{Mplus}. By implementing this model, \citet{Kohli2013PLGC1} examined the development in each stage and estimated a fixed knot for procedural learning task research.

\citet{Preacher2015repara} extended the model with an unknown fixed knot to estimate the knot's variance simultaneously, in which the knot location is allowed to vary and then is viewed as an additional growth factor besides the intercept and two slopes in the latent growth curve (LGC) modeling framework. More importantly, they considered incorporating obesity status for predicting varying knot locations. By re-examining a data set containing longitudinal records of plasma phosphate, they demonstrated the linear-linear functional form with an unknown random knot is useful. However, one drawback of this model lies in that it is fitted in a reparameterized framework. The reparameterized coefficients were no longer directly related to the underlying developmental process and therefore lacked meaningful, substantive interpretation. 

For these reasons, we propose inverse-transformation matrices to derive the growth factor coefficients in the original frame so that they are directly related to the underlying change pattern and interpretable. In addition, we extend the model proposed by \citet{Preacher2015repara} by allowing time-invariant covariates (TICs) to predict individual differences in the initial status and rate-of-change of each stage so that these covariates may account for the heterogeneity in trajectories instead of that only in the knot location. More importantly, we conducted simulation studies to assess how well the Taylor series approximation, which is necessary for reparameterization, works under various conditions.

Last, we propose the model in the framework of individual measurement occasions \citep{Cook1983ITP, Mehta2000people,  Finkel2003cognitive}. This can occur when time is measured precisely (for example, exact date and time of measurement), or if responses are self-initiated or randomly assigned. For instance, in an adolescent smoking study, participants were asked to complete a questionnaire on pocket computers immediately after smoking \citep{Hedeker2006long}. One possible way to fit growth models with individual measurement occasions is the definition variable approach, in which the `definition variables' are defined as observed variables that are employed to adjust model parameters to individual-specific values \citep{Mehta2000people, Mehta2005people}. In our case, these individual-specific values are individual measurement occasions.

The remainder of this article is organized as follows. In the Method section, we first present the model specification of the bilinear spline growth model with time-invariant covariates to estimate a knot with its variability (the full model). Next, we introduce how to reparameterize growth factors and the corresponding path coefficients from the TICs to the growth factors to make them estimable in the SEM framework and transform them back after estimation for interpretation purposes. Following that, we propose one parsimonious model to estimate the knot without variability (the reduced model). We then describe the model estimation as well as the Monte Carlo simulation design for model evaluation. In the Result section, we evaluate the model performance with regard to the non-convergence rate, the proportion of improper solutions, as well as the performance measures, including the relative bias, the empirical standard error (SE), the relative root-mean-squared-error (RMSE) and the empirical coverage probabilities for a nominal $95\%$ confidence interval of each parameter of interest. In the Application section, by applying the proposed model to a longitudinal data set of mathematics achievement scores from the Early Childhood Longitudinal Study, Kindergarten Class of $2010-11$ (ECLS-K: 2011), we give a set of feasible recommendations for real-world practice. Finally, a broad discussion is presented regarding the model's limitations as well as future directions.

\section{Method}\label{Method}
\subsection{Model Specification}\label{M:Specify}
\begin{figure}[ht]
	\centering
	\includegraphics[width=0.5\textwidth]{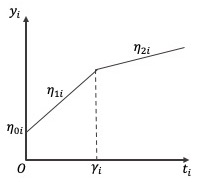}
	\caption{Within-individual Change over Time with Bilinear Spline Functional Form}
	\label{fig:knot}
\end{figure}

This section briefly describes a linear-linear growth curve model with an unknown random knot and time-invariant covariates used to analyze nonlinear change patterns, estimate the knot with its variability, and predict individual differences in trajectories. As shown in Figure \ref{fig:knot}, the linear-linear piecewise latent growth model, which is an extension of LGC, specifies a separate linear function for each of the two stages of the developmental process for each individual. We can express the measure for the $i^{th}$ individual at the $j^{th}$ time point in the framework of individual measurement occasions as
\begin{equation}\label{eq:linear-linear}
y_{ij}=\begin{cases}
\eta_{0i}+\eta_{1i}t_{ij}+\epsilon_{ij} & t_{ij}\le\gamma_{i}\\
\eta_{0i}+\eta_{1i}\gamma_{i}+\eta_{2i}(t_{ij}-\gamma_{i})+\epsilon_{ij} & t_{ij}>\gamma_{i}\\
\end{cases}, 
\end{equation}
where $y_{ij}$ and $t_{ij}$ are the measurement and measurement time of the $i^{th}$ individual at time $j$. In Equation (\ref{eq:linear-linear}), $\eta_{0i}$, $\eta_{1i}$, $\eta_{2i}$ and $\gamma_{i}$ are individual-level intercept, first slope, second slope and knot, which are usually called `growth factors' and all together  determine the functional form of the growth curve of $\boldsymbol{y}_{i}$. 

\figurehere{1}

Note that we cannot fit the linear-linear piecewise change pattern specified in Equation (\ref{eq:linear-linear}) in the LGC framework for two reasons. First, the outcome variable $y_{ij}$ does not have the same model expression pre- and post- knot. Second, the model in Equation (\ref{eq:linear-linear}) specifies a nonlinear relationship between the $y_{ij}$ and the growth factor (or random coefficient in the mixed-effects model framework) $\gamma_{i}$ and then cannot be estimated in the structural equation modeling (SEM) framework directly \citep{Grimm2016growth2}. Following \citet{Tishler1981nonlinear, Seber2003nonlinear, Grimm2016growth, Liu2019knot}, the expression of repeated outcome in Equation (\ref{eq:linear-linear}) before and after the knot can be unified by reparameterization (see Appendix \ref{Supp:1A} for details of reparameterization). We then write the repeated outcome as a linear combination of all four growth factors by the Taylor series expansion \citep{Browne1991Taylor, Grimm2016growth, Liu2019knot} (see Appendix \ref{Supp:1B} for details of Taylor series expansion). Accordingly, the model specified in Equation (\ref{eq:linear-linear}) can be written as a standard LGC model with reparameterized growth factors
\begin{equation}\label{eq:specify_r1}
    \boldsymbol{y}_{i}\approx\boldsymbol{\Lambda}^{'}_{i}\boldsymbol{\eta}^{'}_{i}+\boldsymbol{\epsilon}_{i},
\end{equation}
where $\boldsymbol{y}_{i}$ is a $J\times 1$ vector of the repeated outcomes for the $i^{th}$ person (in which $J$ is the number of measurements), $\boldsymbol{\eta}^{'}_{i}$ is a $4\times 1$ vector of individual-level reparameterized growth factors and $\boldsymbol{\Lambda}^{'}_{i}$, which is a function of time points and the unknown knot, is a $J\times 4$ matrix of factor loadings\footnote{\citet{Preacher2015repara} used similar but slightly different expressions of reparameterized growth factors and corresponding factor loadings.}. Additionally, $\boldsymbol{\epsilon}_{i}$ is a $J\times 1$ vector of residuals for the $i^{th}$ person. For the $i^{th}$ individual, the reparameterized growth factors (the measurement at the knot, the mean of two slopes, the half-difference of two slopes and the knot) and corresponding factor loadings are 
\begin{equation}\label{eq:g_factor_r}
\boldsymbol{\eta}_{i}^{'} = \left(\begin{array}{rrrr}
\eta^{'}_{0i} & \eta^{'}_{1i} & \eta^{'}_{2i} & \delta_{i}
\end{array}\right)^{T}
= \left(\begin{array}{rrrr}
\eta_{0i}+\gamma_{i}\eta_{1i} & \frac{\eta_{1i}+\eta_{2i}}{2} & \frac{\eta_{2i}-\eta_{1i}}{2} & \gamma_{i}-\mu_{\gamma}
\end{array}\right)^{T}
\end{equation}
and
\begin{equation}\label{eq:g_loadings_r}
\begin{aligned}
&\boldsymbol{\Lambda}^{'}_{i} = \left(\begin{array}{rrrr}
1 & t_{ij}-\mu_{\gamma} & |t_{ij}-\mu_{\gamma}| & -\mu^{'}_{\eta_{2}}-\frac{\mu^{'}_{\eta_{2}}(t_{ij}-\mu_{\gamma})}{|t_{ij}-\mu_{\gamma}|}
\end{array}\right)
&(j=1,\cdots, J),
\end{aligned}
\end{equation}
respectively, where $\mu_{\gamma}$ is the knot mean and $\delta_{i}$ is the deviation from the knot mean of the $i^{th}$ individual. The reparameterized growth factors of the $i^{th}$ individual can be further regressed on TICs,
\begin{equation}\label{eq:specify_r2}
    \boldsymbol{\eta}^{'}_{i}=\boldsymbol{\alpha}^{'}+\boldsymbol{B}^{'}\boldsymbol{X}_{i}+\boldsymbol{\zeta}^{'}_{i},
\end{equation}
where $\boldsymbol{\alpha}^{'}$ is a $4\times 1$ vector of reparameterized growth factor intercepts (which is the mean vector of reparameterized growth factors if the TICs are centered), $\boldsymbol{B}^{'}$ is a $4\times c$ matrix of regression coefficients (in which $c$ is the number of TICs) from TICs to reparameterized growth factors, $\boldsymbol{X}_{i}$, which may either be continuous or be binary, is a $c\times 1$ vector of covariates of the $i^{th}$ individual, and $\boldsymbol{\zeta}^{'}_{i}$ is a $4\times 1$ vector of deviations of the $i^{th}$ subject from the reparameterized growth factor means. To simplify estimation, we assume that the (reparameterized) growth factors are normally distributed conditional on individual-level covariates, so $\boldsymbol{\zeta}^{'}_{i}\sim \text{MVN}(\boldsymbol{0}, \boldsymbol{\Psi}^{'}_{\boldsymbol{\eta}})$,
where $\boldsymbol{\Psi}^{'}_{\boldsymbol{\eta}}$ is a $4\times4$ unexplained variance-covariance matrix of reparameterized growth factors. We also assume that individual residuals, $\boldsymbol{\epsilon}_{i}$ in Equation (\ref{eq:specify_r1}), are identical and independent normal distribution over time, that is, $\boldsymbol{\epsilon}_{i}\sim\text{MVN}(\boldsymbol{0}, \theta_{\epsilon}\boldsymbol{I})$, where $\boldsymbol{I}$ is a $J\times J$ identity matrix. Note that the intercepts and coefficients of reparameterized growth factors ($\boldsymbol{\alpha}^{'}$ and $\boldsymbol{B}^{'}$) as well as the unexplained growth factor variances $\boldsymbol{\Psi}^{'}_{\boldsymbol{\eta}}$ are no longer directly related to the underlying developmental process and therefore lack meaningful, substantive interpretation. We demonstrate how to derive the coefficients in the original parameter-space so that all coefficients are interpretable. 

\subsection{Transformation and Inverse-transformation between Two Parameter Spaces}\label{M:Matrix}
When fitting a model, especially a complex model, in the SEM framework, selecting a proper set of initial values can improve the likelihood of convergence and then accelerate the computational process. Generally, descriptive statistics and visualization are tools for researchers to decide suitable initial values for the parameters of interest. However, it may not be straightforward for the reparameterized coefficients. Accordingly, the transformation from parameters in the original frame to those in the reparameterized setting help decide appropriate initial values. More importantly, all reparameterized coefficients are not directly related to the underlying change patterns and then need to be transformed back to be interpretable after estimation. So the inverse-transformation that help derive the coefficients related to the underlying trajectory directly from the estimates are also needed. 

Suppose $\boldsymbol{f}: \mathcal{R}^{4}\rightarrow \mathcal{R}^{4}$ is a function, which takes a point $\boldsymbol{\eta}_{i}\in\mathcal{R}^{4}$ as input and produces the vector $\boldsymbol{f}(\boldsymbol{\eta}_{i})\in\mathcal{R}^{4}$ (i.e., $\boldsymbol{\eta}_{i}^{'}\in\mathcal{R}^{4}$) as output. By the multivariate delta method \citep{Lehmann1998Delta},
\begin{equation}\label{eq:trans_fun}
\boldsymbol{\eta}_{i}^{'}=\boldsymbol{f}(\boldsymbol{\eta}_{i})\sim N\bigg(\boldsymbol{f}(\boldsymbol{\mu_{\eta}}), \boldsymbol{\nabla_{f}}(\boldsymbol{\mu_{\eta}})\boldsymbol{\Psi_{\eta}}\boldsymbol{\nabla_{f}}^{T}(\boldsymbol{\mu_{\eta}})\bigg), 
\end{equation}
where $\boldsymbol{\mu_{\eta}}$ and $\boldsymbol{\Psi_{\eta}}$ are the mean vector and variance-covariance matrix of original growth factors, respectively, and $\boldsymbol{f}$ is defined as
\begin{equation}\nonumber
\boldsymbol{f}(\boldsymbol{\eta}_{i})=\left(\begin{array}{rrrr}
\eta_{0i}+\gamma_{i}\eta_{1i} & \frac{\eta_{1i}+\eta_{2i}}{2} & \frac{\eta_{2i}-\eta_{1i}}{2} &
\gamma_{i}-\mu_{\gamma}
\end{array}\right)^{T}
\end{equation}
as Equation (\ref{eq:g_factor_r}).

Similarly, suppose $\boldsymbol{h}: \mathcal{R}^{4}\rightarrow \mathcal{R}^{4}$ is a function, which takes a point $\boldsymbol{\eta}_{i}^{'}\in\mathcal{R}^{4}$ as input and produces the vector $\boldsymbol{h}(\boldsymbol{\eta}_{i}^{'})\in\mathcal{R}^{4}$ (i.e., $\boldsymbol{\eta}_{i}\in\mathcal{R}^{4}$) as output. By the multivariate delta method,
\begin{equation}\label{eq:inverse_fun}
\boldsymbol{\eta}_{i}=\boldsymbol{h}(\boldsymbol{\eta}_{i}^{'})\sim N\bigg(\boldsymbol{h}(\boldsymbol{\mu}^{'}_{\boldsymbol{\eta}}), \boldsymbol{\nabla_{h}}(\boldsymbol{\mu_{\eta}^{'}})\boldsymbol{\Psi}^{'}_{\boldsymbol{\eta}}\boldsymbol{\nabla_{h}}^{T}(\boldsymbol{\mu_{\eta}^{'}})\bigg), 
\end{equation}
where $\boldsymbol{\mu}^{'}_{\boldsymbol{\eta}}$ and $\boldsymbol{\Psi}^{'}_{\boldsymbol{\eta}}$ are the mean vector and variance-covariance matrix of reparameterized growth factors, respectively, and $\boldsymbol{h}$ is defined as
\begin{equation}\nonumber
\boldsymbol{h}(\boldsymbol{\eta}_{i}^{'})=\left(\begin{array}{rrrr}
\eta^{'}_{0i}-\gamma_{i}\eta^{'}_{1i}+\gamma_{i}\eta^{'}_{2i} & \eta^{'}_{1i}-\eta^{'}_{2i} & \eta^{'}_{1i}+\eta^{'}_{2i} & \delta_{i}+\mu_{\gamma}
\end{array}\right)^{T}.
\end{equation}

Based on Equations (\ref{eq:trans_fun}) and (\ref{eq:inverse_fun}), we can make the transformation between the growth factor means of two parameter spaces by $\boldsymbol{\mu}^{'}_{\boldsymbol{\eta}}
\approx\boldsymbol{f}(\boldsymbol{\mu}_{\boldsymbol{\eta}})$ and $\boldsymbol{\mu_{\eta}}
\approx\boldsymbol{h}(\boldsymbol{\mu}^{'}_{\boldsymbol{\eta}})$, respectively. We can also define the transformation matrix $\boldsymbol{\nabla_{f}}(\boldsymbol{\mu_{\eta}})$ and $\boldsymbol{\nabla_{h}}(\boldsymbol{\mu_{\eta}^{'}})$ between the variance-covariance matrix of two parameter spaces as
\begin{equation}\nonumber
\begin{aligned}
\quad\quad\boldsymbol{\Psi}^{'}_{\boldsymbol{\eta}} 
&\approx \boldsymbol{\nabla_{f}}(\boldsymbol{\mu_{\eta}})\boldsymbol{\Psi_{\eta}}\boldsymbol{\nabla_{f}}^{T}(\boldsymbol{\mu_{\eta}})\\
&=\left(\begin{array}{rrrr}
1 & \mu_{\gamma} & 0 & \mu_{\eta_{1}}\\
0 & 0.5 & 0.5 & 0\\
0 & -0.5 & 0.5 & 0\\
0 & 0 & 0 & 1 
\end{array}\right)\boldsymbol{\Psi_{\eta}}\left(\begin{array}{rrrr}
1 & \mu_{\gamma} & 0 & \mu_{\eta_{1}}\\
0 & 0.5 & 0.5 & 0\\
0 & -0.5 & 0.5 & 0\\
0 & 0 & 0 & 1 
\end{array}\right)^{T}\ \ \\
\end{aligned}
\end{equation}
and
\begin{equation}\nonumber
\begin{aligned}
\boldsymbol{\Psi_{\eta}} &\approx \boldsymbol{\nabla_{h}}(\boldsymbol{\mu_{\eta}^{'}})\boldsymbol{\Psi}^{'}_{\boldsymbol{\eta}}\boldsymbol{\nabla_{h}}^{T}(\boldsymbol{\mu_{\eta}^{'}})\\
&=\left(\begin{array}{rrrr}
1 & -\mu_{\gamma} & \mu_{\gamma} & 0\\0 & 1 & -1 & 0\\0 & 1 & 1 & 0\\0 & 0 & 0 & 1
\end{array}\right) \boldsymbol{\Psi}^{'}_{\boldsymbol{\eta}}\left(\begin{array}{rrrr}
1 & -\mu_{\gamma} & \mu_{\gamma} & 0\\0 & 1 & -1 & 0\\0 & 1 & 1 & 0\\0 & 0 & 0 & 1
\end{array}\right)^{T},
\end{aligned}
\end{equation}
respectively. 

When regressing growth factors on the individual-level covariates as we did in Equation (\ref{eq:specify_r2}), we also need to re-reparameterize the growth factor intercepts and path coefficients. We center the covariates to simplify the calculation. On the one hand, the vector $\boldsymbol{\alpha}^{'}$ is equivalent to the mean vector of the reparameterized growth factor ($\boldsymbol{\mu}^{'}_{\boldsymbol{\eta}}$) when all covariates are centered. Accordingly, we can make the transformation between $\boldsymbol{\alpha}$ and $\boldsymbol{\alpha}^{'}$ by $\boldsymbol{\alpha}^{'}=\boldsymbol{\mu}_{\eta}^{'} \approx\boldsymbol{f}(\boldsymbol{\mu}_{\eta})$ and $\boldsymbol{\alpha}=\boldsymbol{\mu}_{\eta}\approx\boldsymbol{h}(\boldsymbol{\mu}_{\eta}^{'})$. On the other hand, through centering the covariates, we simplify the transformation and inverse-transformation matrices between the original path coefficients ($\boldsymbol{B}$) and the reparameterized path coefficients ($\boldsymbol{B^{'}}$) as $\boldsymbol{\nabla_{f}}(\boldsymbol{\mu_{\eta}})$ and $\boldsymbol{\nabla_{h}}(\boldsymbol{\mu_{\eta}^{'}})$, respectively. The detailed derivation is shown in Appendix \ref{Supp:1D}. 

When fitting the model using software such as the \textit{R} package \textit{OpenMx}, which allows for matrix calculation on those estimates from the model by the function \textit{mxAlgebra()} \citep{User2018OpenMx}, we only need to provide $\boldsymbol{h}(\boldsymbol{\mu_{\eta}^{'}})$ and $\boldsymbol{\nabla_{h}}(\boldsymbol{\mu_{\eta}^{'}})$, and the package can derive the coefficients in the original setting and generate the point estimates with their standard errors of the growth factor coefficients in the original setting automatically. Other SEM software such as \textit{Mplus} can also calculate new parameters to be derived from those estimated automatically by specifying their relationship in the \textit{NEW} command. However, we need to derive the expression for each cell of the mean vector and the variance-covariance matrix of the original growth factors since \textit{Mplus} does not allow for matrix algebra. All these expressions are in Appendix \ref{Supp:1E}.

\subsection{Model Estimation}\label{M:Estimate}
For the $i^{th}$ individual, the expected mean vector and the variance-covariance structure of the repeated measurements of the model given in Equations (\ref{eq:specify_r1}) and (\ref{eq:specify_r2}) can be expressed as
\begin{equation}\label{eq:mean_r}
\boldsymbol{\mu}_{i}=\boldsymbol{\Lambda}^{'}_{i}(\boldsymbol{\alpha}^{'}+\boldsymbol{B}^{'}\boldsymbol{\mu_{X})}
\end{equation}
and
\begin{equation}\label{eq:var_r}
\boldsymbol{\Sigma}_{i}=\boldsymbol{\Lambda}^{'}_{i}\boldsymbol{\Psi}^{'}_{\boldsymbol{\eta}}\boldsymbol{\Lambda}_{i}^{'T}+\boldsymbol{\Lambda}^{'}_{i}\boldsymbol{B^{'}\Phi B}^{'T}\boldsymbol{\Lambda}^{'T}_{i}+\theta_{\epsilon}\boldsymbol{I},
\end{equation}
where $\boldsymbol{\mu_{X}}$ and $\boldsymbol{\Phi}$ are the mean vector ($c\times 1$) and the variance-covariance matrix ($c\times c$) of the TICs, respectively.

The parameters in the model given in Equations (\ref{eq:specify_r1}) and (\ref{eq:specify_r2}) include the mean vector and unexplained variance-covariance matrix of reparameterized growth factors, the re-expressed path coefficients as well as the means and the variance-covariance structure of covariates. Then we can obtain the parameters in the original setting, as shown in Section \ref{M:Matrix}. $\boldsymbol{\Theta}_{1}$ and $\boldsymbol{\Theta}_{1}^{'}$ shown below
\begin{equation}\nonumber
\begin{aligned}
\boldsymbol{\Theta}_{1}&=\{\boldsymbol{\alpha}, \boldsymbol{\Psi_{\eta}}, \boldsymbol{B}, \boldsymbol{\mu_{X}}, \boldsymbol{\Phi}, \theta_{\epsilon}\}\\
&=\{\mu_{\eta_{0}}, \mu_{\eta_{1}}, \mu_{\eta_{2}}, \mu_{\gamma}, \psi_{00}, \psi_{01}, \psi_{02}, \psi_{0\gamma}, \psi_{11}, \psi_{12}, \psi_{1\gamma}, \psi_{22}, \psi_{2\gamma}, \psi_{\gamma\gamma}, \boldsymbol{B}, \boldsymbol{\mu_{X}}, \boldsymbol{\Phi}, \theta_{\epsilon}\}
\end{aligned}
\end{equation}
and
\begin{equation}\nonumber
\begin{aligned}
\boldsymbol{\Theta}_{1}^{'}&=\{\boldsymbol{\alpha}^{'}, \boldsymbol{\Psi}^{'}_{\boldsymbol{\eta}}, \boldsymbol{B}^{'}, \boldsymbol{\mu_{X}}, \boldsymbol{\Phi}, \theta_{\epsilon}\}\\
&=\{\mu_{\eta_{0}}^{'}, \mu_{\eta_{1}}^{'}, \mu_{\eta_{2}}^{'}, \mu_{\gamma}, \psi_{00}^{'}, \psi_{01}^{'}, \psi_{02}^{'}, \psi_{0\gamma}^{'}, \psi_{11}^{'}, \psi_{12}^{'}, \psi_{1\gamma}^{'}, \psi_{22}^{'}, \psi_{2\gamma}^{'}, \psi_{\gamma\gamma}, \boldsymbol{B}^{'}, \boldsymbol{\mu_{X}}, \boldsymbol{\Phi}, \theta_{\epsilon}\}
\end{aligned}
\end{equation}
list the parameters in the original and reparameterized frames, respectively. 

We then use full information maximum likelihood (FIML) to estimate $\boldsymbol{\Theta}_{1}^{'}$ due to the potential heterogeneity of individual contributions to the likelihood. The log-likelihood function of each individual and that of the overall sample can be expressed as
\begin{equation}\label{eq:loglik1}
\log lik_{i}(\boldsymbol{\Theta}_{1}^{'}|\boldsymbol{y}_{i})=C-\frac{1}{2}\ln|\boldsymbol{\Sigma}_{i}|-\frac{1}{2}\big(\boldsymbol{y}_{i}-\boldsymbol{\mu}_{i})^{T}\boldsymbol{\Sigma}_{i}^{-1}(\boldsymbol{y}_{i}-\boldsymbol{\mu}_{i}\big),
\end{equation}
and
\begin{equation}\label{eq:loglik2}
\log lik(\boldsymbol{\Theta}_{1}^{'})=\sum_{i=1}^{n}\log lik_{i}(\boldsymbol{\Theta}_{1}^{'}|\boldsymbol{y}_{i}),
\end{equation}
respectively, where $C$ is a constant, $n$ is the number of individuals, $\boldsymbol{\mu}_{i}$ and $\boldsymbol{\Sigma}_{i}$ are the mean vector and the variance-covariance matrix of $\boldsymbol{y}_{i}$, which have been defined in Equations (\ref{eq:mean_r}) and (\ref{eq:var_r}), respectively. We construct the full model using the R package \textit{OpenMx} with the optimizer CSOLNP \citep{OpenMx2016package, Pritikin2015OpenMx, Hunter2018OpenMx, User2018OpenMx}, with which we are able to fit the model and obtain interpretable coefficients efficiently, as shown in Section \ref{M:Matrix}.

\subsection{Reduced Model}\label{M:Reduced}
With an assumption that the change-point is roughly similar for each individual, we can fix the between-individual differences in the knot to $0$ and construct a reduced model to estimate a knot without considering variability. We can specify the reduced model as
\begin{align}
&\boldsymbol{y}_{i}=\boldsymbol{\Lambda}^{'}_{i}\boldsymbol{\eta}^{'}_{i}+\boldsymbol{\epsilon}_{i},\nonumber\\
&\boldsymbol{\eta}^{'}_{i}=\boldsymbol{\alpha}^{'}+\boldsymbol{B}^{'}\boldsymbol{X}_{i}+\boldsymbol{\zeta}^{'}_{i},\nonumber
\end{align}
where $\boldsymbol{\eta}^{'}_{i}$ is a $3\times1$ vector of reparameterized growth factor (the measurement of the knot, the mean of two slopes, and the half-difference of two slopes), and $\boldsymbol{\Lambda}_{i}$, a function of time points and a fixed change-point, is a $J\times 3$ matrix of factor loadings. We can further express the reparameterized growth factors and corresponding factor loadings as
\begin{equation}\nonumber
\boldsymbol{\eta}_{i}^{'} = \left(\begin{array}{rrr}
\eta^{'}_{0i} & \eta^{'}_{1i} & \eta^{'}_{2i} 
\end{array}\right)^{T}
= \left(\begin{array}{rrr}
\eta_{0i}+\gamma\eta_{1i} & \frac{\eta_{1i}+\eta_{2i}}{2} & \frac{\eta_{2i}-\eta_{1i}}{2} 
\end{array}\right)^{T}
\end{equation}
and
\begin{equation}\nonumber
\begin{aligned}
&\boldsymbol{\Lambda}^{'}_{i} = \left(\begin{array}{rrr}
1 & t_{ij}-\gamma & |t_{ij}-\gamma| 
\end{array}\right)
&(j=1,\cdots, J).
\end{aligned}
\end{equation}
The growth factor intercepts $\boldsymbol{\alpha}^{'}$, path coefficients $\boldsymbol{B}^{'}$ and the deviation $\boldsymbol{\zeta}^{'}_{i}$ of the $i^{th}$ individual from the reparameterized growth factor means also reduce to be a $3\times1$ vector, $3\times c$ matrix and $3\times1$ vector, respectively. The transformation and inverse-transformation functions are also reduced accordingly. Specifically, we only need the first three entries of the functions $\boldsymbol{f}$ and $\boldsymbol{h}$, and then the first three rows and the first three columns in the matrices $\boldsymbol{\nabla_{f}}(\boldsymbol{\mu_{\eta}})$ and $\boldsymbol{\nabla_{h}}(\boldsymbol{\mu_{\eta}^{'}})$, since only three growth factors need to be reparameterized. 

For the $i^{th}$ individual, the expected mean vector and the variance-covariance matrix of the repeated outcomes of this model are 
\begin{equation}\label{eq:mean_f}
\boldsymbol{\mu}_{i}=\boldsymbol{\Lambda}^{'}_{i}(\boldsymbol{\alpha}^{'}+\boldsymbol{B}^{'}\boldsymbol{\mu_{X}})
\end{equation}
and
\begin{equation}\label{eq:var_f}
\boldsymbol{\Sigma}_{i}=\boldsymbol{\Lambda}^{'}_{i}\boldsymbol{\Psi}^{'}_{\boldsymbol{\eta}}\boldsymbol{\Lambda}_{i}^{'T}+\boldsymbol{\Lambda}^{'}_{i}\boldsymbol{B}^{'}\boldsymbol{\Phi B}^{'T}\boldsymbol{\Lambda}^{'T}_{i}+\theta_{\epsilon}\boldsymbol{I},
\end{equation}
respectively. For this reduced model, $\boldsymbol{\Theta}_{2}$ and $\boldsymbol{\Theta}_{2}^{'}$ are defined as
\begin{equation}\nonumber
\begin{aligned}
\boldsymbol{\Theta}_{2}&=\{\boldsymbol{\alpha}, \gamma, \boldsymbol{\Psi_{\eta}}, \boldsymbol{B}, \boldsymbol{\mu_{X}}, \boldsymbol{\Phi}, \theta_{\epsilon}\}\\
&=\{\mu_{\eta_{0}}, \mu_{\eta_{1}}, \mu_{\eta_{2}}, \gamma, \psi_{00}, \psi_{01}, \psi_{02},  \psi_{11}, \psi_{12},  \psi_{22},  \boldsymbol{B}, \boldsymbol{\mu_{X}}, \boldsymbol{\Phi}, \theta_{\epsilon}\}\ \ \ 
\end{aligned}
\end{equation}
and
\begin{equation}\nonumber
\begin{aligned}
\boldsymbol{\Theta}_{2}^{'}&=\{\boldsymbol{\alpha}^{'}, \gamma, \boldsymbol{\Psi}^{'}_{\boldsymbol{\eta}}, \boldsymbol{B}^{'}, \boldsymbol{\mu_{X}}, \boldsymbol{\Phi}, \theta_{\epsilon}\}\\
&=\{\mu_{\eta_{0}}^{'}, \mu_{\eta_{1}}^{'}, \mu_{\eta_{2}}^{'}, \gamma, \psi_{00}^{'}, \psi_{01}^{'}, \psi_{02}^{'},  \psi_{11}^{'}, \psi_{12}^{'},  \psi_{22}^{'}, \boldsymbol{B}^{'}, \boldsymbol{\mu_{X}}, \boldsymbol{\Phi}, \theta_{\epsilon}\},
\end{aligned}
\end{equation}
and they list the parameters in the original and reparameterized setting, respectively. By replacing $\boldsymbol{\Theta}_{1}^{'}$ in Equations (\ref{eq:loglik1}) and (\ref{eq:loglik2}) with $\boldsymbol{\Theta}_{2}^{'}$ and updating $\boldsymbol{\mu}_{i}$ and $\boldsymbol{\Sigma}_{i}$ as such defined in Equations (\ref{eq:mean_f}) and (\ref{eq:var_f}), we have the likelihood function of each individual and that of the overall sample. We build the reduced model using the R package \textit{OpenMx} with the optimizer CSOLNP and employ the FIML technique to estimate the parameters. We provide the \textit{OpenMx} syntax for the full model and its reduced version as well as a demonstration in the online appendix (\url{https://github.com/Veronica0206/Dissertation_projects}). For the researchers who are interested in using \textit{Mplus}, we also provide \textit{Mplus} 8 syntax for both models in the online appendix. 

\section{Model Evaluation}\label{M:Evaluate}
The proposed method is evaluated using a Monte Carlo simulation study with two goals. The first goal is to evaluate how the approximation based on the Taylor series expansion affects the performance measures, including the relative bias, the empirical standard error (SE), the relative root-mean-square error (RMSE), and the empirical coverage probability for a nominal $95\%$ confidence interval of each parameter. We provide the definitions and estimates of these four performance metrics in Table \ref{tbl:metric}. The second goal is to examine how the reduced model performed sufficiently well as compared to the full model.

\tablehere{1}

In the simulation study, we followed \citet{Morris2019simulation} and decided the number of replications  $S=1,000$ by an empirical approach. We conducted a pilot simulation run and found that standard errors of all coefficients except the unexplained intercept variance (i.e., $\psi_{00}$) were less than $0.15$. As the most important performance metric in the simulation study was the (relative) bias, and since we wanted to keep its Monte Carlo standard error\footnote{$\text{Monte Carlo SE(Bias)}=\sqrt{Var(\hat{\theta})/S}$ \citep{Morris2019simulation}.} lower than $0.005$, we needed at least $900$ replications. We then decided to proceed with $S=1,000$ for more conservative consideration. 

\subsection{Design of Simulation Study}\label{Simu:design}
As mentioned earlier, the parameters of the most interest in the full model are the knot, its variance and the path coefficients to the knot. The conditions hypothesized to influence the estimation of these knot parameters, along with other model parameters, include the number of individuals, the number of repeated measurements, the knot locations, the knot variances, shapes of trajectories, measurement precision, and the proportion of growth factor variances explained by time-invariant covariates. Accordingly, we did not investigate some conditions that presumably would not affect the model performance meaningfully. For instance, the mean vector and variance-covariance matrix of the first three growth factors (i.e., the intercept and two slopes) usually change with the measurement scale and time scale in practice. As shown in Table \ref{tbl:simu_design}, accordingly, we fixed the mean of intercept and first slope and only changed that of the second slope. We also fixed the variance-covariance matrix of the first three growth factors and kept the index of dispersion ($\sigma^{2}/\mu$) of each at a tenth scale, guided by previous simulation studies for growth models \citep{Bauer2003GMM, Kohli2011PLGC, Kohli2015PLGC2}. Further, the growth factors were set to be positively correlated to a moderate degree ($\rho=0.3$).

\tablehere{2}

We list all conditions that we considered in the simulation design in Table \ref{tbl:simu_design}. For a model to analyze longitudinal data, the factor that we are most interested in is the number of repeated measures. Generally, the model should perform better if the study duration is longer, which we wanted to investigate by the simulation study. Another important factor for the linear-linear trajectory is the knot location. Intuitively, the model should perform the best if the knot is in the middle of study duration; it is our interest to test this conjecture as well. To this end, we chose two different levels of the number of measurements: $6$ and $10$. We selected $6$ as the minimum number of repeated outcomes to make the full model fully identified\footnote{Although the information of model identification of linear-linear growth model with an unknown knot is scarce, it has been proved that the bilinear latent growth model can be identified with at least five waves with a specified knot at the halfway of study duration \citep{Bollen2005LCM}.}. For the conditions with $6$ repeated measures, we set the knot at halfway of study duration ($\mu_{\gamma}=2.5$). The other level $10$ was considered for two reasons. First, we wanted to evaluate whether a more extended study duration would improve model performance. Second, $10$ measurements allowed us to place knots at different locations, say the halfway knot ($\mu_{\gamma}=4.5$) as well as a left-shifted knot ($\mu_{\gamma}=3.5$) and right-shifted knot ($\mu_{\gamma}=5.5$) from the middle point of study duration. Around each wave, we allowed a time-window $(-0.25, +0.25)$, which is a `medium' deviation as in \citet{Coulombe2015ignoring}, for individual measurement occasions. 

In order to fit the full model in the SEM framework and obtain interpretable coefficients, the approximation was introduced by the Taylor series expansion. As shown in Appendix \ref{Supp:1B}, one necessary condition to apply this approximation is that the knot variance approaches to $0$. We then investigated $3$ levels of magnitude of between-person differences in the knot to evaluate how the introduced approximation affects the model performance and to compare the full model versus the reduced model. We set the knot standard deviation as $0$, $0.3$ and $0.6$ as the zero, medium and large difference in the knot location. Note that the magnitude of the knot standard deviation is relative to the scale of measurement occasions. 

Additionally, we examined how the shape of change patterns, which was quantified by the standardized difference between two slopes, affects the full model. It is noted that we considered both positive and negative differences ($\eta_{1i}<\eta_{2i}$ and $\eta_{1i}>\eta_{2i}$, respectively) to assess model performance in the two cases as shown in Figure \ref{fig:proj1_2cases}. For each case, we considered $3$ levels of standardized difference between two slopes to evaluate the models. We also considered $2$ levels of effects of exogenous variables on endogenous growth factors\footnote{We set up TICs to explain moderate ($13\%$) and substantial ($26\%$) variances of such factors \citep{Cohen1988R}.}, $2$ levels of sample size, and $2$ levels of measurement precision as shown in Table \ref{tbl:simu_design}. 

\subsection{Data Generation and Simulation Step}\label{evaluation:step}
For each condition which is listed in Table \ref{tbl:simu_design}, the general steps of the simulation study for the models were outlined as follows:
\begin{enumerate}
	\item Generated data for growth factors and TICs simultaneously using the R package \textit{MASS} \citep{Venables2002Statistics} (details are provided in Appendix \ref{Supp:1F}),
	\item Generated the time structure with $J$ scaled and equally-spaced waves $t_{j}$ and obtained individual measurement occasions: $t_{ij}\sim U(t_{j}-\Delta, t_{j}+\Delta)$ by allowing disturbances around each wave,
	\item Calculated factor loadings, which are functions of individual measurement occasions and the knot, for each individual,
	\item Calculated values of the repeated measurements based on growth factors, factor loadings, and residual variances.
	\item Implemented the full model and the reduced model on simulated data, estimated the parameters, constructed corresponding 95\% Wald CIs,
	\item Repeated the above steps until after obtaining $1,000$ convergent solutions.
\end{enumerate}

\section{Results}\label{Result}
\subsection{Model Convergence and Proper Solution}\label{Preliminary}
In this section, we first examined the convergence rate and component fit measures of each condition. The convergence\footnote{In our project, convergence is defined as to achieve \textit{OpenMx} status code $0$, which suggests a successful optimization, until up to $10$ attempts with different sets of starting values \citep{OpenMx2016package}.} rate of each condition was investigated. Based on our simulation studies, the full model and its reduced version converged satisfactorily (the convergence rate of the full model achieved at least $95\%$ while that of the reduced model was $100\%$). Out of a total of $576$ conditions, $490$ conditions reported $100\%$ convergence rate and $67$ conditions reported convergence rate of $99\%$ to $100\%$. The worst scenario in terms of the non-convergence rate across all conditions was $46/1046$, indicating that we needed to repeat the steps described in Section \ref{evaluation:step} $1,046$ times to have $1,000$ replications with a convergent solution. It occurred under the condition with the smaller sample size (i.e., $n=200$), the shorter study duration (i.e., $J=6$), less precise measurements (i.e., $\theta_{\epsilon}=2$), the largest knot standard deviation (i.e., $sd(\gamma)=0.6$), and the smallest positive standardized difference between two slopes (i.e., $d_{z}=1$). 

We also conducted diagnostics to investigate improper solutions, such as estimates of growth factor variances less than $0$, and/or estimates of correlations between growth factors beyond $[-1, 1]$. Table \ref{tbl:Improper} demonstrates the number of improper solutions yielded by the full model under conditions with $10$ repeated measures, the knot midway through the total time of observation, and $13\%$ explained variance of the latent growth factors. The improper solutions include negative knot variances and its out-of-range (i.e., out of $[-1, 1]$) correlations with any other growth factors. When the population value of the knot variance was $0$ or relatively small (i.e., $sd(\gamma)=0.3$), the number of improper solutions was relatively large. 

\tablehere{3}

For the scenarios including knots with variability, we observed that the proper solution rate has positive associations with the magnitude of standardized difference between two slopes, the measurement precision, the sample size, and the knot standard deviation. The sign of $d_{z}$, the shifted knot locations, or the proportion of explained variance of growth factors only affected the number of improper solutions slightly, although reduced follow-up time inflated this number. When such improper solutions occurred, we replaced the full model with its reduced version for the model evaluation.

\subsection{Performance Measures}\label{Primary}
In this section, we present simulation results in terms of performance measures, which include the relative bias, empirical SE, relative RMSE and empirical coverage probability for a nominal $95\%$ confidence interval for each parameter of interest. In general, the full model estimates coefficients unbiasedly, precisely, with satisfactory confidence interval coverage. We first provide the summary statistics (specifically, median and range) for each performance metric of each parameter across the conditions with $10$ repeated measurements and the midway knot. Then we discuss how the simulation conditions affect these performance metrics.  

Tables \ref{tbl:rBias} and \ref{tbl:empSE} present the median (range) of the relative bias and empirical SE, respectively, for each parameter of interest across all conditions with $10$ repeated measurements and the halfway knot for the full model and its reduced version. We first calculated the relative bias/empirical SE of each parameter across $1,000$ replications under each condition and then summarized the relative biases/empirical SEs of each parameter over all aforementioned conditions using the median (range) of relative bias/empirical SE. As shown in Tables \ref{tbl:rBias} and \ref{tbl:empSE}, both models generated unbiased point estimates with small empirical SEs to a large extent, and the full model performed better than its reduced version as the ranges of relative biases of the parameters from the full model were narrower than those from the reduced model. Specifically, for the full model, the magnitude of relative biases of the growth factor means was under $0.03$, that of path coefficients to the intercept and two slopes was under $0.12$, and that of unexplained variances of the intercept and two slopes was under $0.07$. From Table \ref{tbl:rBias}, the full model may produce biased estimates for the knot variances and the path coefficients to the knot: the median values of relative biases of $\beta_{1\gamma}$, $\beta_{2\gamma}$ and $\psi_{\gamma\gamma}$ were $-0.2712$, $-0.2709$ and $0.0403$, respectively. 

\tablehere{4}

\tablehere{5}

\begin{figure}[ht]
	\centering
	\includegraphics[width=1.0\textwidth]{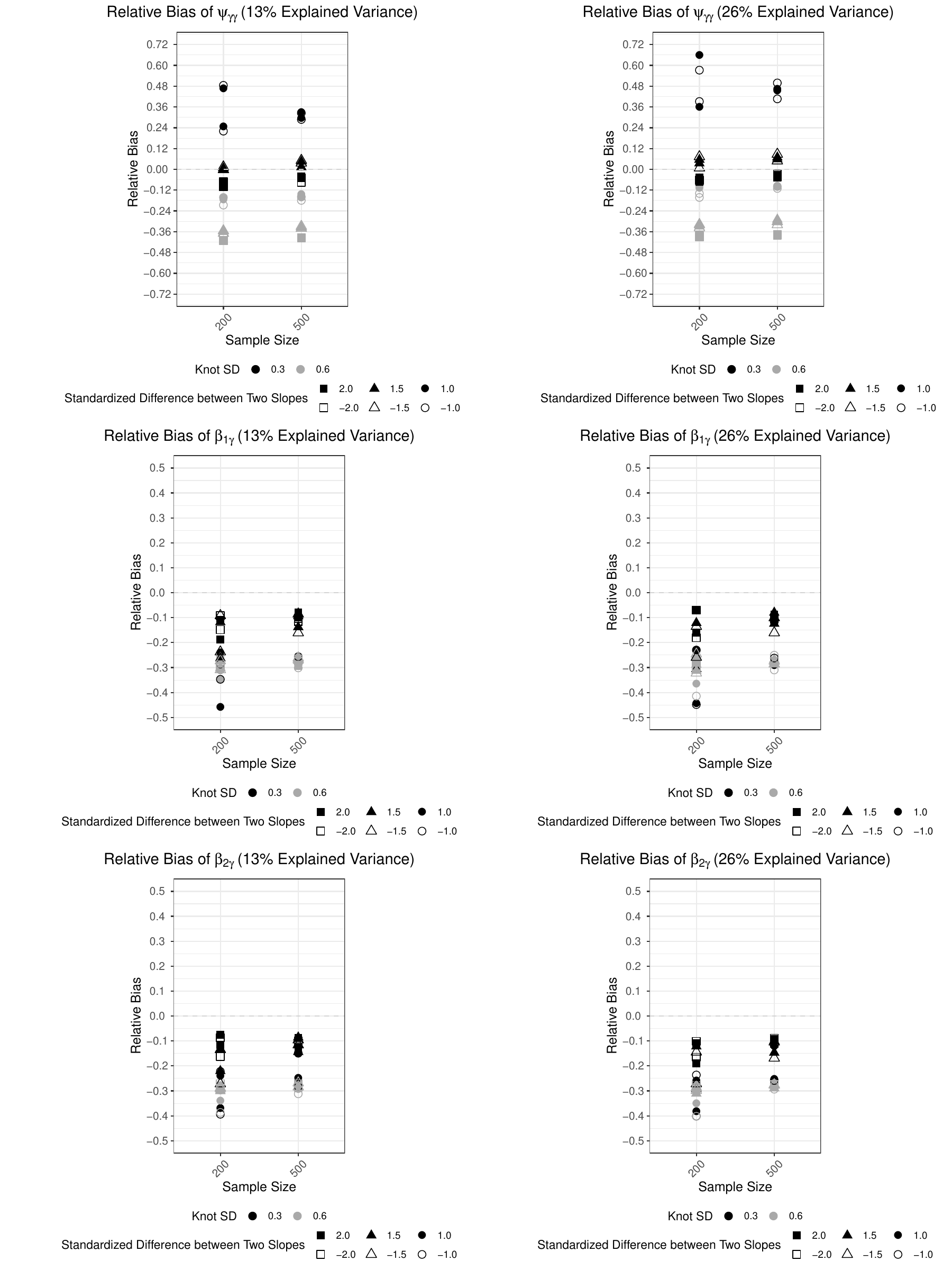}
	\caption{Relative Biases of Path Coefficients and Residuals of Knot under Conditions with $10$ Repeated Measures \& Midway Knot}
	\label{fig:rBiasknot}
\end{figure}

We plotted the relative bias under each condition with $10$ repeated measures and the midway knot for $\psi_{\gamma\gamma}$, $\beta_{1\gamma}$ and $\beta_{2\gamma}$ in Figure \ref{fig:rBiasknot} to further investigate the relative bias pattern for the knot variance and the path coefficients to the knot. From the figure, we observed how the conditions we set up in the simulation design affected these estimates. First, under the conditions where the knot standard deviation was $0.3$ and/or the standardized difference between two slopes was medium or large (i.e., $d_{z}=1.5$ or $d_{z}=2.0$), relative biases were smaller. Second, the unexplained knot variance, $\psi_{\gamma\gamma}$, was underestimated under the conditions with $0.6$ knot standard deviation while could be overestimated under the conditions with $0.3$ knot standard deviation, and the path coefficients to the knot (i.e., $\beta_{1\gamma}$ and $\beta_{2\gamma}$) were underestimated under all investigated conditions. Third, other conditions, such as the sample size or the proportion of growth factor variance explained by the covariates, only influenced the relative biases slightly. 

\figurehere{2}

From Table \ref{tbl:empSE}, estimates from the full model and its reduced version were precise: the magnitude of empirical standard errors of slope or knot parameters were around $0.10$, although those values of intercept parameters were relatively large: the empirical standard error of the intercept mean, and that of the unexplained intercept variance achieved $0.35$ and $2.35$, respectively. 
	
Table \ref{tbl:rRMSE} lists the median (range) of relative RMSE of each parameter for both models under the conditions with $10$ repeated measures and the midway knot, which combines bias and precision to examine the point estimates holistically. From the table, both models estimated parameters accurately. The magnitude of relative RMSEs of growth factor means was under $0.05$, and the median RMSE value of path coefficients to the intercept or slopes was around $0.20$. 
	
\tablehere{6}
	
Table \ref{tbl:CP} presents the median (range) of the coverage probability (CP) for the full model and its reduced version under the conditions with $10$ repeated measurements and the midway knot. Under the majority of conditions, the full model performed well in terms of empirical coverage as the median values of coverage probabilities of all parameters were around $0.95$. We noticed that the coverage probabilities of knot variance and path coefficients to the knot could be conservative under the conditions with $0.6$ knot standard deviation. 
	
\tablehere{7}
	
In summary, based on our simulation study, the full model performed better than the reduced model in terms of performance measures and generated unbiased estimates with small variance and proper target coverage probability. As pointed out earlier, one necessary condition for applying the Taylor series approximation is that the knot variance approaches $0$. Through the simulation study, we found that the full model was robust under the conditions with the medium level of knot standard deviation (i.e., $\text{sd}(\gamma)=0.3$ for the scaled measurement occasions) though was less satisfactory when the knot standard deviation was large, which is within our expectation to see. Factors, including sample size and measurement precision, also influenced model performance. Specifically, the larger sample size (i.e., $n=500$) and more precise measurement ($\theta_{\epsilon}=1$) improved performance measures. The trajectory shape quantified by the standardized difference between two slopes affected the performance measures shown in Figure \ref{fig:rBiasknot}. We also examined the model performance beyond conditions summarized in Tables \ref{tbl:rBias}, \ref{tbl:empSE}, \ref{tbl:rRMSE} and \ref{tbl:CP} and found that shorter study duration (i.e., $J=6$) and left- (right-) shifted knot locations only affected the performance measures slightly. 

\section{Application}
This section demonstrates the use of the full model to approximate nonlinear trajectories and estimate a knot that varies across individuals. This application has two goals. The first goal is to compare the fitness of bilinear spline functional form to three common parametric change patterns: linear, quadratic and Jenss-Bayley growth curve. The second goal is to show how to select the model between the full model and the reduced version. We randomly selected $400$ students from the Early Childhood Longitudinal Study Kindergarten Cohort: 2010-11 (ECLS-K: 2011) with complete records of repeated mathematics IRT scaled scores, demographic information (sex, race, and age at each wave), school information (baseline school location and baseline school type), and social-economic status (baseline family income and the highest education level between parents)\footnote{The total sample size of ECLS-K: 2011 is $n=18174$. The number of entries after removing records with missing values (i.e., rows with any of NaN/-9/-8/-7/-1) is $n=2290$.}. 

ECLS-K: 2011 is a nationally representative longitudinal sample of US children enrolled in about $900$ kindergarten programs starting from $2010-2011$ school year. Children's mathematics ability was assessed in nine waves: fall and spring of kindergarten ($2010-2011$), first ($2011-2012$) and second ($2012-2013$) grade, respectively, as well as spring of $3^{rd}$ ($2014$), $4^{th}$ ($2015$) and $5^{th}$ ($2016$), respectively. Only about $30\%$ students were sampled in the fall of $2011$ and $2012$ \citep{Le2011ECLS}. Noting that two possible time scales are available in the data set: students' age and their grade-in-school, we decided to use children's age (in months) to have individual measurement occasions in the main analysis and conduct a sensitivity analysis where students' grade-in-school was used as the time scale. In the subsample, $51\%$ of children were male, and $49\%$ of children were female. Additionally, $44\%$ of students were White, $5\%$ were Black, $38\%$ were Hispanic, $8\%$ were Asian, and $5\%$ were others. For this analysis, we dichotomized the variable race to be White ($44\%$) and others ($56\%$). At the start of the study, $11\%$ and $89\%$ students were in private and public school, respectively. We treated the variables school location (ranged between $1$ and $4$), highest parents' education (ranged between $0$ and $8$) and family income (ranged between $1$ and $18$) as continuous variables, and corresponding mean (SD) was $2.06\ (1.00)$, $5.18\ (2.10)$ and $11.56\ (5.42)$, respectively.

\subsection*{Main Analysis}
\begin{figure}[ht]
	\centering
	\includegraphics[width=1.0\textwidth]{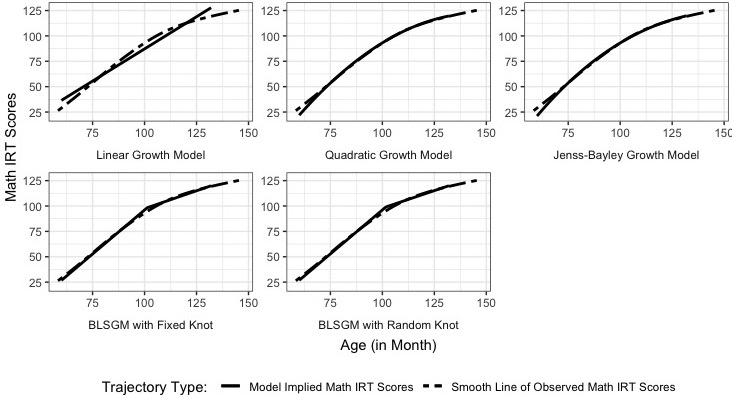}
	\caption{Model Implied Trajectory and Smooth Line of Observed mathematics IRT Scores}
	\label{fig:est_lines}
\end{figure}

In this section, we use children's age (in months) as the time metric to have individual measurement occasions. We first fit the full and reduced linear-linear piecewise models as well as three parametric growth curve functions, i.e., linear, quadratic and Jenss-Bayley. The graphical representations of the model implied curves together with the smooth line of the observed mathematics IRT scores are shown in Figure \ref{fig:est_lines}. From the figure, the nonlinear functional forms fit better than the linear function generally; and the models with bilinear spline change patterns fit better than the parametric nonlinear models to the first two or three measurement occasions. As shown in Figure \ref{fig:est_lines}, both the quadratic growth curve and the Jenss-Bayley tended to underestimate the mathematics achievement in the early stage. 

\figurehere{3}

Table \ref{tbl:info} provides the obtained estimated likelihood, information criteria (including AIC and BIC), residuals of each latent growth curve model. From the table, the model with the linear-linear functional form and a random knot has the largest estimated likelihood, smallest AIC and smallest residual though its BIC is slightly larger than the LGC with the quadratic function. Moreover, from Table \ref{tbl:info}, the full model consistently performs better than its reduced version for this empirical example, indicating that we need to consider the variability of the knot, although the model implied trajectory of both captures the overall observed data well as shown in Figure \ref{fig:est_lines}. We then fit the full model to investigate the impact of covariates, such as sex, race, parents' highest education (baseline), family income (baseline), school type (baseline), and school location (baseline), on the knot variance (and other growth factor variances). Its model fit information is also provided in Table \ref{tbl:info}.

\tablehere{8}

Table \ref{tbl:est} presents the estimates of parameters of interest. It is noticed that post-knot development in mathematics skills slowed down substantially (on average $1.731$ and $0.688$ per month in the pre- and post-knot stage, respectively). The knot was estimated at $102$ month on average ($8.5$-year old). We noted that the initial mathematics performance (60-month old in our case) was positively associated with family income, private school, and parents' highest education. Moreover, the family income also can explain the individual difference in the development rate of mathematics IRT scores in the first stage. Further, boys improved more rapidly than girls in terms of mathematics ability in the first stage. Additionally, students coming from schools in more urban areas arrived at the change-point earlier. 

\tablehere{9}

\subsection*{Sensitivity Analysis}
In this section, we use students' grade-in-school as the time scale, where the time metric takes on discrete values (for example, $0$ is for the kindergarten fall semester, and $0.5$ in years or $6$ in months is for the kindergarten spring semester). We list the estimated likelihood, AIC, BIC, and the residual variance of the linear-linear piecewise growth model in Table \ref{tbl:info}\footnote{Note that the model constructed in the sensitivity analysis cannot be compared with that in the main analysis directly since the raw data sets used in the two parts were different.}. Upon further examination, the mathematics ability slowed down between the sixth and the seventh wave (around the fall semester of Grade $3$), matching the estimate in the main analysis: students in Grade $3$ usually aged between $8$- to $9$-year old. The effect size of the estimates of the path coefficients to the knot increased slightly, but the direction of the estimates stayed the same. 

\section{Discussion}\label{Discussion}
In this article, we extended an existing linear-linear piecewise growth model with an unknown random knot to the framework of individual measurement occasions to account for the heterogeneity of the measurement time. We explored the predictors to explain the between-individual differences in the within-individual change patterns. We also proposed using the multivariate delta method to derive the mean vector and variance-covariance matrix of growth factors as well as the path coefficients directly related to the underlying change patterns.  More importantly, we examined the proposed method by extensive simulation studies to investigate whether the approximation introduced by the Taylor series expansion affects the model performance.

To demonstrate how well the introduced approximation works, we compared the full model (that requires the Taylor series approximation) with the reduced model and found that the full model performed better than the reduced one in terms of the performance measures under the conditions with varying knot locations. The simulation study also showed that the full model performed well under the conditions with a medium level knot standard deviation, which aligns with the essential condition using the Taylor series expansion. Based on the result of simulation studies, we recommend using the bilinear spline growth model with an unknown random knot if the assumption of the same knot location is indefensible. In the scenario where \textit{a priori} knowledge regarding the knot variability is scarce, we recommend fitting the full model and the reduced model and then select the `best' one as we did in the Application section. Additionally, based on the simulation studies, the reduced model, which only estimates the knot itself, is a good backup if the full model fails to converge or generates improper solutions or the knot variance is out of research interest, although sometimes at a little cost of a small increase in bias.

The (inverse-) transformation between growth factors in two parameter-spaces are valuable for the reparameterized longitudinal model. One drawback of the reparameterized longitudinal model initially was that the reparameterized coefficients were no longer related to the underlying developmental process and therefore lacked meaningful, substantive interpretation, thus requiring a back transformation of the model parameters. In the linear-linear piecewise growth model with an unknown random knot, this transformation is at individual-level as each individual has a set of `personal' growth factors. We then proposed to use the multivariate delta method to simplify the calculation. The matrix form is especially convenient when we use the \textit{R} package \textit{OpenMx} that can carry out matrix multiplication automatically. The transformation functions and matrices for other reparameterized longitudinal models can also be proposed.

We demonstrated the use of the proposed method on a subset with $n=400$ from ECLS-K: 2011 and provided a set of recommendations for possible issues that researchers may face in a real-world data analysis. First of all, we recommend selecting a proper functional form to depict trajectories before identifying the possible time-invariant covariates to explain the between-individual difference in the within-individual change, though this selection is not straightforward. Many factors, including but not limited to the research question, the fit between the model implied and the observed trajectory, and the information criteria, drive this selection process. It is worth considering either the full or the reduced model to estimate the knot if it is part of the research interest. If the interest lies in the fitting of nonlinear trajectories, it may be appropriate to fit the change patterns with several functional forms and then select the `best' one. It is noted that the model fit information, sometimes, fails to lead unequivocal selection as in our example. Although either the full model or the quadratic functional forms can be viewed as an `acceptable' model, the full model is considered better if it is important to capture children's mathematics achievement in the early stage of the study. The selection between the full and the reduced model can only rely on the information criteria such as the BIC as they usually fit the trajectories equally well, as shown in Figure \ref{fig:est_lines}. To demonstrate the influence of the time scale has on the estimates, we conducted a sensitivity analysis in the application section. The fixed effects of growth factors obtained from the model with different time scales were almost the same, though the estimated effect size of path coefficients increased when we used the grade-in-school as the time scale. This difference is not surprising as students in the same grade-in-school could be of different ages. In practice, we recommend selecting the time scale by research questions instead of based on statistical criteria. 

To summarize, in this article, we focused on the growth model with a bilinear spline functional form because it is the most straightforward, intuitive and useful. It can be generalized to a linear spline with multiple knots or a nonlinear spline (such as a linear-polynomial piecewise), where the corresponding (inverse-) transformation matrices can also be derived. Besides, it is feasible to extend the current work to address longitudinal studies with dropout under the assumption of missing at random thanks to the FIML technique. Additionally, we focused on time-invariant covariates in the current work, which is helpful to identify baseline characteristics to explain the variability of growth factors. It can also be extended to investigate time-varying covariates to explain the individual-difference in the trajectories.

\vspace{\fill}\pagebreak

\appendix
\renewcommand{\theequation}{A.\arabic{equation}}
\setcounter{equation}{0}

\renewcommand{\thesection}{Appendix \Alph{section}}
\renewcommand{\thesubsection}{A.\arabic{subsection}}

\section{Formula Derivation}\label{Supp:1}
\subsection{The Reparameterizing Procedure for Growth Factors}\label{Supp:1A}
\citet{Tishler1981nonlinear} and \citet{Seber2003nonlinear} showed that the bilinear spline regression model can be written as either the minimum or maximum response value of two trajectories. By extending such expressions to the LGC framework, two forms of bilinear spline for the $i^{th}$ individual are shown in Figure \ref{fig:proj1_2cases}. In the left panel ($\eta_{1i}>\eta_{2i}$), the measurement $y_{ij}$ should always be the minimum value of two lines; that is, $y_{ij}=\min{(\eta_{0i}+\eta_{1i}t_{ij}, \eta_{02i}+\eta_{2i}t_{ij})}$. To unify the expression of measurements pre- and post-knot, we have the following equation
\begin{equation}\label{eq:left}
\begin{aligned}
y_{ij} &= \min{(\eta_{0i} + \eta_{1i}t_{ij}, \eta_{02i} + \eta_{2i}t_{ij})}\\
&= \frac{1}{2}\big(\eta_{0i} + \eta_{1i}t_{ij} + \eta_{02i} + \eta_{2i}t_{ij} - 
|\eta_{0i} + \eta_{1i}t_{ij} - \eta_{02i} - \eta_{2i}t_{ij}|\big)\\
&= \frac{1}{2}\big(\eta_{0i} + \eta_{1i}t_{ij} + \eta_{02i} + \eta_{2i}t_{ij}\big) - 
\frac{1}{2}\big(|\eta_{0i} + \eta_{1i}t_{ij} - \eta_{02i} - \eta_{2i}t_{ij}|\big)\\
&= \frac{1}{2}\big(\eta_{0i} + \eta_{02i} + \eta_{1i}t_{ij} + \eta_{2i}t_{ij}\big) - 
\frac{1}{2}\big(\eta_{1i} - \eta_{2i}\big)|t_{ij} - \gamma_{i}|\\
&= \eta^{'}_{0i} + \eta^{'}_{1i}\big(t_{ij}-\gamma_{i}\big) + \eta^{'}_{2i}|t_{ij} - \gamma_{i}|\\
&= \eta^{'}_{0i} + \eta^{'}_{1i}\big(t_{ij}-\gamma_{i}\big) + \eta^{'}_{2i}\sqrt{(t_{ij} - \gamma_{i})^2},
\end{aligned}
\end{equation}
where $\eta^{'}_{0i}$, $\eta^{'}_{1i}$, $\eta^{'}_{2i}$ and $\gamma_{i}$ are the measurement at the knot, mean of two slopes, the half-difference between two slopes and knot for the $i^{th}$ individual. Through straightforward algebra, the measurement $y_{ij}$ of the bilinear spline in the right panel, in which the measurement $y_{ij}$ should always be the maximum value of two lines, has the identical final form as Equation (\ref{eq:left}).

\figurehere{A.1}

\subsection{Taylor Series Expansion}\label{Supp:1B}
For the $i^{th}$ individual, suppose we define a function $f(\gamma_{i})$ and its first derivative with respect to $\gamma_{i}$\footnote{It is noted that we have $4$ degree-of-freedom when expressing the repeated outcomes of a linear combination of $4$ growth factors \citep{Harring2006nonlinear}. To simplify subsequent calculation, we viewed $\eta_{0i}$, $\eta_{1i}$, $\eta_{2i}$ and $\gamma_{i}$ (i.e., the growth factors in the original setting) as independent variables. It is also noted that $\gamma_{i}$ is intact during the process of reparameterization \citep{Preacher2012repara}.}, shown below
\begin{equation}\nonumber
f(\gamma_{i})=\eta^{'}_{0i} + \eta^{'}_{1i}\big(t_{ij}-\gamma_{i}\big) + \eta^{'}_{2i}\sqrt{(t_{ij} - \gamma_{i})^2} 
\end{equation}
and
\begin{equation}\nonumber
f^{'}(\gamma_{i})=\eta_{1i}-\eta^{'}_{1i}-\frac{\eta^{'}_{2i}(t_{ij}-\gamma_{i})}{\sqrt{(t_{ij}-\gamma_{i})^2}}=-\eta^{'}_{2i}-\frac{\eta^{'}_{2i}(t_{ij}-\gamma_{i})}{\sqrt{(t_{ij}-\gamma_{i})^2}},
\end{equation}
respectively. Then the Taylor series expansion of $f(\gamma_{i})$ can be expressed as
\begin{equation}\nonumber
\begin{aligned}
f(\gamma_{i})&= f(\mu_{\gamma})+\frac{f'(\mu_{\gamma})}{1!}(\gamma_{i}-\mu_{\gamma})+\cdots\\
&= \eta^{'}_{0i}+\eta^{'}_{1i}(t_{ij}-\mu_{\gamma})+\eta^{'}_{2i}\sqrt{(t_{ij} - \mu_{\gamma})^2} +
(\gamma_{i}-\mu_{\gamma})\bigg[-\eta^{'}_{2i}-\frac{\eta^{'}_{2i}(t_{ij}-\mu_{\gamma})}{|t_{ij}-\mu_{\gamma}|}\bigg]+\cdots\\
&\approx \eta^{'}_{0i}+\eta^{'}_{1i}(t_{ij}-\mu_{\gamma})+\eta^{'}_{2i}|t_{ij}-\mu_{\gamma}| +
(\gamma_{i}-\mu_{\gamma})\bigg[-\mu^{'}_{\eta_{2}}-\frac{\mu^{'}_{\eta_{2}}(t_{ij}-\mu_{\gamma})}{|t_{ij}-\mu_{\gamma}|}\bigg],
\end{aligned}
\end{equation}
from which we then have the reparameterized growth factors and the corresponding factor loadings for the $i^{th}$ individual.

\subsection{Details of the transformations between intercept coefficients and path coefficients}\label{Supp:1D}
\begin{equation}
\begin{aligned}
&\boldsymbol{\mu}^{'}_{\boldsymbol{\eta}}
\approx\boldsymbol{f}(\boldsymbol{\mu}_{\boldsymbol{\eta}})
\Longleftrightarrow E(\boldsymbol{\alpha}^{'}+\boldsymbol{B^{'}X}_{i}+\boldsymbol{\zeta}^{'}_{i})\approx\boldsymbol{f}(E(\boldsymbol{\alpha}+\boldsymbol{BX}_{i}+\boldsymbol{\zeta}_{i}))
\Longleftrightarrow 
\boldsymbol{\alpha}^{'}
\approx\boldsymbol{f}(\boldsymbol{\alpha})\nonumber\\
&\boldsymbol{\mu_{\eta}}
\approx\boldsymbol{h}(\boldsymbol{\mu}^{'}_{\boldsymbol{\eta}})
\Longleftrightarrow E(\boldsymbol{\alpha}+\boldsymbol{BX}_{i}+\boldsymbol{\zeta}_{i})\approx\boldsymbol{h}(E(\boldsymbol{\alpha}^{'}+\boldsymbol{B^{'}X}_{i}+\boldsymbol{\zeta}^{'}_{i}))
\Longleftrightarrow \boldsymbol{\alpha}
\approx\boldsymbol{h}(\boldsymbol{\alpha}^{'})\\
&\boldsymbol{\Psi_{\eta}^{'}}\approx
\boldsymbol{\nabla_{f}}(\boldsymbol{\mu_{\eta}})\boldsymbol{\Psi_{\eta}}\boldsymbol{\nabla_{f}}^{T}(\boldsymbol{\mu_{\eta}})\\ &\Longleftrightarrow Var(\boldsymbol{\alpha}^{'}+\boldsymbol{B^{'}X}_{i}+\boldsymbol{\zeta}_{i}^{'})\approx \boldsymbol{\nabla_{f}}(\boldsymbol{\mu_{\eta}})Var(\boldsymbol{\alpha}+\boldsymbol{BX}_{i}+\boldsymbol{\zeta}_{i})\boldsymbol{\nabla_{f}}^{T}(\boldsymbol{\mu_{\eta}})\\
&\Longleftrightarrow Var(\boldsymbol{B^{'}X}_{i}+\boldsymbol{\zeta}_{i}^{'})\approx\boldsymbol{\nabla_{f}}(\boldsymbol{\mu_{\eta}})Var(\boldsymbol{BX}_{i}+\boldsymbol{\zeta}_{i})\boldsymbol{\nabla_{f}}^{T}(\boldsymbol{\mu_{\eta}})\\
&\Longleftrightarrow \boldsymbol{B}^{'}Var(\boldsymbol{X}_{i})\boldsymbol{B}^{'T}+Var(\boldsymbol{\zeta}_{i}^{'})\approx\boldsymbol{\nabla_{f}}(\boldsymbol{\mu_{\eta}})\boldsymbol{B}Var(\boldsymbol{X}_{i})\boldsymbol{B}^{T}\boldsymbol{\nabla_{f}}^{T}(\boldsymbol{\mu_{\eta}})+\boldsymbol{\nabla_{f}}(\boldsymbol{\mu_{\eta}})Var(\boldsymbol{\zeta}_{i})\boldsymbol{\nabla_{f}}^{T}(\boldsymbol{\mu_{\eta}})\ \ \ \ \ \ \ \ \ \ \ \ \ \\
&\Longleftrightarrow \boldsymbol{B}^{'}\approx\boldsymbol{\nabla_{f}}(\boldsymbol{\mu_{\eta}})\boldsymbol{B}\\
&\boldsymbol{\Psi_{\eta}}\approx
\boldsymbol{\nabla_{h}}(\boldsymbol{\mu_{\eta}^{'}})\boldsymbol{\Psi}^{'}_{\boldsymbol{\eta}}\boldsymbol{\nabla_{h}}^{T}(\boldsymbol{\mu_{\eta}^{'}})\\ &\Longleftrightarrow Var(\boldsymbol{\alpha}+\boldsymbol{BX}_{i}+\boldsymbol{\zeta}_{i})\approx \boldsymbol{\nabla_{h}}(\boldsymbol{\mu_{\eta}^{'}})Var(\boldsymbol{\alpha}^{'}+\boldsymbol{B^{'}X}_{i}+\boldsymbol{\zeta}^{'}_{i})\boldsymbol{\nabla_{h}}^{T}(\boldsymbol{\mu_{\eta}^{'}})\\
&\Longleftrightarrow Var(\boldsymbol{BX}_{i}+\boldsymbol{\zeta}_{i})\approx\boldsymbol{\nabla_{h}}(\boldsymbol{\mu_{\eta}^{'}})Var(\boldsymbol{B^{'}X}_{i}+\boldsymbol{\zeta}^{'}_{i})\boldsymbol{\nabla_{h}}^{T}(\boldsymbol{\mu_{\eta}^{'}})\nonumber\\
&\Longleftrightarrow \boldsymbol{B}Var(\boldsymbol{X}_{i})\boldsymbol{B}^{T}+Var(\boldsymbol{\zeta}_{i})\approx\boldsymbol{\nabla_{h}}(\boldsymbol{\mu_{\eta}^{'}})\boldsymbol{B}^{'}Var(\boldsymbol{X}_{i})\boldsymbol{B}^{'T}\boldsymbol{\nabla_{h}}^{T}(\boldsymbol{\mu_{\eta}^{'}})+\boldsymbol{\nabla_{h}}(\boldsymbol{\mu_{\eta}^{'}})Var(\boldsymbol{\zeta}^{'}_{i})\boldsymbol{\nabla_{h}}^{T}(\boldsymbol{\mu_{\eta}^{'}})\\
&\Longleftrightarrow \boldsymbol{B}\approx\boldsymbol{\nabla_{h}}(\boldsymbol{\mu_{\eta}^{'}})\boldsymbol{B}^{'}\nonumber
\end{aligned}
\end{equation}

\subsection{Expression of each cell of the re-reparameterized mean vector and variance-covariance matrix}\label{Supp:1E}
\begin{equation}
\begin{aligned}
\mu_{\eta_{0}}&\approx\mu^{'}_{\eta_{0}}-\mu_{\gamma}\mu^{'}_{\eta_{1}}+\mu_{\gamma}\mu^{'}_{\eta_{2}}\\
\mu_{\eta_{1}}&=\mu^{'}_{\eta_{1}}-\mu^{'}_{\eta_{2}}\\
\mu_{\eta_{2}}&=\mu^{'}_{\eta_{2}}+\mu^{'}_{\eta_{1}}\\
\mu_{\gamma}&=\mu_{\gamma}\\
\psi_{00}&\approx(\psi_{11}^{'}+\psi_{22}^{'}-2\psi_{12}^{'})\mu_{\gamma}^{2}+2(\psi_{02}^{'}-\psi_{01}^{'})\mu_{\gamma}+\psi_{00}^{'}\nonumber\\
\psi_{01}&\approx(2\psi_{12}^{'}-\psi_{11}^{'}-\psi_{22}^{'})\mu_{\gamma}+(\psi_{01}^{'}-\psi_{02}^{'})\nonumber\\
\psi_{02}&\approx(\psi_{22}^{'}-\psi_{11}^{'})\mu_{\gamma}+(\psi_{01}^{'}+\psi_{02}^{'})\nonumber\\
\psi_{0\gamma}&\approx(\psi_{2\gamma}^{'}-\psi_{1\gamma}^{'})\mu_{\gamma}+\psi_{0\gamma}^{'}\nonumber\\
\psi_{11}&=\psi_{11}^{'}+\psi_{22}^{'}-2\psi_{12}^{'}\nonumber\\
\psi_{12}&=\psi_{11}^{'}-\psi_{22}^{'}\nonumber\\
\psi_{1\gamma}&=\psi_{1\gamma}^{'}-\psi_{2\gamma}^{'}\nonumber\\
\psi_{22}&=\psi_{11}^{'}+\psi_{22}^{'}+2\psi_{12}^{'}\nonumber\\
\psi_{2\gamma}&=\psi_{1\gamma}^{'}+\psi_{2\gamma}^{'}\nonumber\\
\psi_{\gamma\gamma}&=\psi_{\gamma\gamma}^{'}\nonumber
\end{aligned}
\end{equation}

\subsection{Generate Exogenous Variables and Growth Factors Simultaneously}\label{Supp:1F}
To generate growth factors and exogenous variables simultaneously as a multivariate normal distribution, we need to specify the mean vector and the variance-covariance matrix of the distribution. In our setting, the mean vector can be represented in equation
\begin{equation}\nonumber
\boldsymbol{\mu}=\left(\begin{array}{rrr}
\boldsymbol{\alpha} & 0 & 0
\end{array}\right),
\end{equation}
where $\boldsymbol{\alpha}$ and $\begin{array}{rr}(0 & 0)\end{array}$ are the mean vector of the growth factors and that of the TICs, respectively. Moreover, through straightforward linear algebra, the underlying variance-covariance matrix of the distribution is
\begin{equation}\nonumber
\boldsymbol{\Sigma}=\left(\begin{array}{rr}
\boldsymbol{B\Phi B+\Psi_{\eta}} & \boldsymbol{B\Phi} \\
\boldsymbol{B\Phi} & \boldsymbol{\Phi} 
\end{array}\right),
\end{equation}
where $\boldsymbol{B}$ is the matrix of path coefficients, $\boldsymbol{\Psi_{\eta}}$ and $\boldsymbol{\Phi }$ are the variance-covariance matrices of the growth factors and the TICs, respectively. Accordingly, we can generate the growth factors and TICs simultaneously, which follows a multivariate normal distribution $N_{6}(\boldsymbol{\mu}, \boldsymbol{\Sigma})$.


\bibliographystyle{apalike}
\bibliography{Paper1}


\newpage
\begin{table}
\centering
\begin{threeparttable}
\caption{Performance Metric: Definitions and Estimates}
\begin{tabular}{p{4cm}p{4.5cm}p{5.5cm}}
	\hline
	\hline
	\textbf{Criteria} & \textbf{Definition} & \textbf{Estimate} \\
	\hline
	Relative Bias & $E_{\hat{\theta}}(\hat{\theta}-\theta)/\theta$ & $\sum_{s=1}^{S}(\hat{\theta}-\theta)/S\theta$ \\
	Empirical SE & $\sqrt{Var(\hat{\theta})}$ & $\sqrt{\sum_{s=1}^{S}(\hat{\theta}-\bar{\theta})^{2}/(S-1)}$ \\
	Relative RMSE & $\sqrt{E_{\hat{\theta}}(\hat{\theta}-\theta)^{2}}/\theta$ & $\sqrt{\sum_{s=1}^{S}(\hat{\theta}-\theta)^{2}/S}/\theta$ \\
	Coverage Probability & $Pr(\hat{\theta}_{\text{low}}\le\theta\le\hat{\theta}_{\text{upper}})$ & $\sum_{s=1}^{S}I(\hat{\theta}_{\text{low},s}\le\theta\le\hat{\theta}_{\text{upper},s})/S$\\
	\hline
	\hline
\end{tabular}
\label{tbl:metric}
\begin{tablenotes}
\small
\item[1] {$\theta$: the population value of the parameter of interest} \\
\item[2] {$\hat{\theta}$: the estimate of $\theta$} \\
\item[3] {$S$: the number of replications and set as $1,000$ in our simulation study} \\
\item[4] {$s=1,\dots,S$: indexes the replications of the simulation} \\
\item[5] {$\hat{\theta}_{s}$: the estimate of $\theta$ from the $s^{th}$ replication} \\
\item[6] {$\bar{\theta}$: the mean of $\hat{\theta}_{s}$'s across replications} \\
\item[7] {$I()$: an indicator function}
\end{tablenotes}
\end{threeparttable}
\end{table}

\begin{table}[ht]
	\centering
	\begin{threeparttable}
	\setlength{\tabcolsep}{5pt}
	\renewcommand{\arraystretch}{0.8}
	\caption{Simulation Design for the Models}
	\begin{tabular}{p{5.5cm} p{9.8cm}}
	\hline
	\hline
	\multicolumn{2}{c}{\textbf{Fixed Conditions}}\\
	\hline
	\textbf{Variables} & \textbf{Conditions} \\
	\hline
	Mean of the Intercept & $\mu_{\eta_{0}}=100$ \\
	Variance of the Intercept & $\psi_{00}=25$ \\
	Mean of the 1st Slope & $\mu_{\eta_{1}}=-5$ \\
	Variance of Slopes & $\psi_{11}=\psi_{22}=1$ \\
	Correlations of Growth Factors & $\rho=0.3$ \\
	\hline
	\hline
	\multicolumn{2}{c}{\textbf{Manipulated Conditions}}\\
	\hline
	\multicolumn{2}{l}{\textbf{Partially Crossed Design}}\\
	\hline
	\textbf{Variables} & \textbf{Conditions} \\
	\hline
	\multirow{2}{*}{Time ($t_{j}$)} & $6$ scaled and equally spaced $(j=0 \cdots, J-1, J=6)$\\
	& $10$ scaled and equally spaced $(j=0, \cdots, J-1, J=10)$\\
	\hline
	\multirow{2}{*}{(Mean of) the knot} & $\mu_{\gamma}$ at $t=2.5$ for $J=6$\\
	& $\mu_{\gamma}$ at $t=3.5$ or $4.5$ or $5.5$ for $J=10$\\
	\hline
	Individual $t_{ij}$ & $t_{ij} \sim U(t_{j}-\Delta, t_{j}+\Delta) (j=0, \cdots, J-1; \Delta=0.25)$ \\
	\hline
	\multirow{2}{*}{Sample Size} & $n=200$ \\
	& $n=500$ \\
	\hline
	\multicolumn{2}{l}{\textbf{Full Factorial Design}}\\
	\hline
	\textbf{Variables} & \textbf{Conditions} \\
	\hline
	\multirow{3}{*}{Standardized Difference $d_{z}$} 
	& $\mu_{\eta_{1}}-\mu_{\eta_{2}}=\pm{1.6}$ ($1.0-$unit Standardized Difference) \\
	& $\mu_{\eta_{1}}-\mu_{\eta_{2}}=\pm{2.4}$ ($1.5-$unit Standardized Difference)\\
	& $\mu_{\eta_{1}}-\mu_{\eta_{2}}=\pm{3.2}$ ($2.0-$unit Standardized Difference)\\
	\hline
	\multirow{3}{*}{Variance of Knots} 
	& $\psi_{\gamma\gamma}=0$ ($sd(\gamma)/(t_{j}-t_{j-1})=0$) \\
	& $\psi_{\gamma\gamma}=0.09$ ($sd(\gamma)/(t_{j}-t_{j-1})=0.3$) \\
	& $\psi_{\gamma\gamma}=0.36$ ($sd(\gamma)/(t_{j}-t_{j-1})=0.6$) \\
	\hline
	\multirow{2}{*}{Coefficients ($\boldsymbol{B}$)} & TICs explain $13\%$ variability of growth factors \\
	& TICs explain $26\%$ variability of growth factors \\
	\hline
	\multirow{2}{*}{Residual Variance} & $\theta_{\epsilon}=1$ \\
	& $\theta_{\epsilon}=2$ \\
	\hline
	\hline
	\end{tabular}
	\label{tbl:simu_design}
	\end{threeparttable}
\end{table}

\begin{table}[ht]
	\centering
	\begin{threeparttable}
	\setlength{\tabcolsep}{5pt}
	\renewcommand{\arraystretch}{0.8}
	\caption{Number of Improper Solutions among $1,000$ Convergent Replications under Conditions with $10$ Repeated Measures \& Midway Knot, $13\%$ Explained Variance}
	\begin{tabular}{p{2.2cm}|p{2.0cm}|p{2.5cm}|ll|ll}
	\hline
	& & & \multicolumn{2}{c}{$\theta_{\epsilon}=1$} & \multicolumn{2}{c}{$\theta_{\epsilon}=2$}\\ 
	\hline
	& & & \textbf{$n=200$} & \textbf{$n=500$} & \textbf{$n=200$} & \textbf{$n=500$}\\
	\hline
	\multirow{6}{*}{Small $d_{z}$}
	& \multirow{3}{*}{Positive $d_{z}$} 
	& $sd(\gamma)=0$ & $555//48$\tnote{1} & $516//22$ & $550//40$ & $507//29$ \\
	& & $sd(\gamma)=0.3$ & $117//35$ & $7//11$ & $278//48$ & $168//21$ \\
	& & $sd(\gamma)=0.6$ & $4//3$ & $0//0$ & $58//25$ & $7//6$ \\
	\cline{2-7}
	&\multirow{3}{*}{Negative $d_{z}$} 
	& $sd(\gamma)=0$ & $535//37$ & $542//27$ & $568//44$ & $531//27$ \\
	& & $sd(\gamma)=0.3$ & $106//49$ & $24//14$ & $279//59$ & $134//39$ \\
	& & $sd(\gamma)=0.6$ & $0//8$ & $0//0$ & $67//45$ & $8//17$ \\
	\hline
	\multirow{6}{*}{Medium $d_{z}$} 
	&\multirow{3}{*}{Positive $d_{z}$} 
	& $sd(\gamma)=0$ & $563//49$ & $553//28$ & $583//39$ & $573//22$ \\
	& & $sd(\gamma)=0.3$ & $17//14$ & $1//0$ & $141//27$ & $40//11$ \\
	& & $sd(\gamma)=0.6$ & $0//1$ & $0//0$ & $9//2$ & $0//0$ \\
	\cline{2-7}
	& \multirow{3}{*}{Negative $d_{z}$} 
	& $sd(\gamma)=0$ & $568//42$ & $542//26$ & $576//50$ & $513//32$ \\
	& & $sd(\gamma)=0.3$ & $19//17$ & $1//0$ & $130//64$ & $22//32$ \\
	& & $sd(\gamma)=0.6$ & $0//0$ & $0//0$ & $4//23$ & $0//0$ \\
	\hline
	\multirow{6}{*}{Large $d_{z}$} 
	&\multirow{3}{*}{Positive $d_{z}$} 
	& $sd(\gamma)=0$ & $555//44$ & $555//36$ & $562//45$ & $557//19$ \\
	& & $sd(\gamma)=0.3$ &  $1//0$ & $0//0$ & $27//19$ & $0//0$ \\
	& & $sd(\gamma)=0.6$ & $0//0$ & $0//0$ & $0//0$ & $0//0$ \\
	\cline{2-7}
	&\multirow{3}{*}{Negative $d_{z}$} 
	& $sd(\gamma)=0$ & $548//35$ & $514//30$ & $555//36$ & $523//33$ \\
	& & $sd(\gamma)=0.3$ & $1//2$ & $0//0$ & $37//39$ & $2//6$ \\
	& & $sd(\gamma)=0.6$ & $0//0$ & $0//0$ & $0//2$ & $0//0$ \\
	\hline
	\end{tabular}
	\label{tbl:Improper}
	\begin{tablenotes}
	\small
	\item[1] {$555//48$ indicates that among $1,000$ convergent replications, we have $555$ and $48$ improper solutions due to negative knot variances and out-of-range correlations of the knot with other growth factors, respectively.}
	\end{tablenotes}
	\end{threeparttable}
\end{table}

\begin{table}[ht]
	\centering
	\begin{threeparttable}
	\setlength{\tabcolsep}{5pt}
	\renewcommand{\arraystretch}{0.8}
	\caption{Median (Range) of Relative Bias of Each Parameter Across Conditions with $10$ Repeated Measures \& Midway Knot}
	\begin{tabular}{llrr}
	\hline
	\hline
	& \textbf{Para.} & \textbf{Reduced Model} & \textbf{Full Model} \\
	\hline
	& & Median (Range) & Median (Range) \\
	\hline
	\hline
	\multirow{4}{*}{\textbf{Mean Vector}} 
	& $\mu_{\eta_{0}}$ & $0.0000$ ($-0.0005$, $0.0006$) & $0.0000$ ($-0.0006$, $0.0006$) \\
	& $\mu_{\eta_{1}}$ & $0.0000$ ($-0.0098$, $0.0100$) & $0.0000$ ($-0.0099$, $0.0100$) \\
	& $\mu_{\eta_{2}}$ & $0.0000$ ($-0.0063$, $0.0280$) & $0.0000$ ($-0.0062$, $0.0273$) \\
	& $\mu_{\gamma}$ & $0.0000$ ($-0.0014$, $0.0009$) & $0.0000$ ($-0.0017$, $0.0006$) \\
	\hline
	\hline
	\multirow{4}{*}{\textbf{Path Coef. of $x_{1}$}} 
	& $\beta_{10}$ & $0.0030$ ($-0.0823$, $0.0773$) & $0.0027$ ($-0.0345$, $0.0367$) \\
	& $\beta_{11}$ & $0.0023$ ($-0.3069$, $0.3055$) & $0.0010$ ($-0.1113$, $0.1052$) \\
	& $\beta_{12}$ & $0.0016$ ($-0.2984$, $0.2970$) & $0.0002$ ($-0.0940$, $0.0946$) \\
	& $\beta_{1\gamma}$ & ---\tnote{1} & $-0.2712$ (NA\tnote{2}, NA) \\
	\hline
	\hline
	\multirow{4}{*}{\textbf{Path Coef. of $x_{2}$}} 
	& $\beta_{20}$ & $0.0006$ ($-0.0697$, $0.0690$) & $0.0007$ ($-0.0216$, $0.0250$) \\
	& $\beta_{21}$ & $0.0006$ ($-0.2901$, $0.2950$) & $0.0000$ ($-0.0882$, $0.0933$) \\
	& $\beta_{22}$ & $-0.0017$ ($-0.2888$, $0.2901$) & $-0.0019$ ($-0.0901$, $0.0878$) \\
	& $\beta_{2\gamma}$ & --- & $-0.2709$ (NA, NA) \\
	\hline
	\hline
	\multirow{4}{*}{\textbf{Unexplained Variance}} 
	& $\psi_{00}$ & $-0.0107$ ($-0.0409$, $0.0212$) & $-0.0107$ ($-0.0235$, $0.0018$) \\
	& $\psi_{11}$ & $-0.0070$ ($-0.0753$, $0.1692$) & $-0.0138$ ($-0.0657$, $0.0182$) \\
	& $\psi_{22}$ & $-0.0080$ ($-0.0703$, $0.1638$) & $-0.0130$ ($-0.0647$, $0.0132$) \\
	& $\psi_{\gamma\gamma}$ & --- & $0.0403$ ($-0.4151$, NA) \\
	\hline
	\hline
	\end{tabular}
	\label{tbl:rBias}
	\begin{tablenotes}
	\small
	\item[1] {--- indicates that the relative biases are not available from the reduced model.}
	\item[2] {NA indicates that the bounds of relative bias is not available. The model performance under the conditions with $0$ population value of difference in knot is of interest where the relative bias of those knot coefficients would go infinity.} \\
	\end{tablenotes}
	\end{threeparttable}
\end{table}

\begin{table}[ht]
	\centering
	\begin{threeparttable}
	\setlength{\tabcolsep}{5pt}
	\renewcommand{\arraystretch}{0.8}
	\caption{Median (Range) of Empirical Standard Error of Each Parameter Across Conditions with $10$ Repeated Measures \& Midway Knot}
	\begin{tabular}{llrr}
	\hline
	\hline
	& \textbf{Para.} & \textbf{Reduced Model} & \textbf{Full Model} \\
	\hline
	& & Median (Range) & Median (Range) \\
	\hline
	\hline
	\multirow{4}{*}{\textbf{Mean Vector}} 
	& $\mu_{\eta_{0}}$ & $0.2576$ ($0.1903$, $0.3550$) & $0.2578$ ($0.1903$, $0.3547$) \\
	& $\mu_{\eta_{1}}$ & $0.0548$ ($0.0387$, $0.0755$) & $0.0548$ ($0.0387$, $0.0755$) \\
	& $\mu_{\eta_{2}}$ & $0.0552$ ($0.0400$, $0.0755$) & $0.0552$ ($0.0400$, $0.0762$) \\
	& $\mu_{\gamma}$ & $0.0418$ ($0.0173$, $0.0854$) & $0.0412$ ($0.0173$, $0.0922$) \\
	\hline
	\hline
	\multirow{4}{*}{\textbf{Path Coef. of $x_{1}$}} 
	& $\beta_{10}$ & $0.2761$ ($0.1957$, $0.3734$) & $0.2768$ ($0.1967$, $0.3735$) \\
	& $\beta_{11}$ & $0.0579$ ($0.0412$, $0.0800$) & $0.0583$ ($0.0412$, $0.0806$) \\
	& $\beta_{12}$ & $0.0574$ ($0.0412$, $0.0819$) & $0.0583$ ($0.0412$, $0.0819$) \\
	& $\beta_{1\gamma}$ & ---\tnote{1} & $0.0387$ ($0.0100$, $0.0990$) \\
	\hline
	\hline
	\multirow{4}{*}{\textbf{Path Coef. of $x_{2}$}} 
	& $\beta_{20}$ & $0.2725$ ($0.1980$, $0.3744$) & $0.2720$ ($0.1980$, $0.3751$) \\
	& $\beta_{21}$ & $0.0574$ ($0.0412$, $0.0806$) & $0.0583$ ($0.0412$, $0.0806$) \\
	& $\beta_{22}$ & $0.0578$ ($0.0412$, $0.0800$) & $0.0582$ ($0.0412$, $0.0812$) \\
	& $\beta_{2\gamma}$ & --- & $0.0400$ ($0.0100$, $0.1054$) \\
	\hline
	\hline
	\multirow{4}{*}{\textbf{Unexplained Variance}} 
	& $\psi_{00}$ & $1.6912$ ($1.1734$, $2.3690$) & $1.6902$ ($1.1741$, $2.3501$) \\
	& $\psi_{11}$ & $0.0748$ ($0.0490$, $0.1122$) & $0.0747$ ($0.0490$, $0.1095$) \\
	& $\psi_{22}$ & $0.0741$ ($0.0490$, $0.1140$) & $0.0740$ ($0.0510$, $0.1091$) \\
	& $\psi_{\gamma\gamma}$ & --- & $0.0387$ ($0.0000$, $0.1732$) \\
	\hline
	\hline
	\end{tabular}
	\label{tbl:empSE}
	\begin{tablenotes}
	\small
	\item[1] {--- indicates that the empirical standard errors are not available from the reduced model.}
	\end{tablenotes}
	\end{threeparttable}
\end{table}

\begin{table}[ht]
	\centering
	\begin{threeparttable}
	\setlength{\tabcolsep}{5pt}
	\renewcommand{\arraystretch}{0.8}
	\caption{Median (Range) of Relative RMSE of Each Parameter Across Conditions with $10$ Repeated Measures \& Midway Knot}
	\begin{tabular}{llrr}
	\hline
	\hline
	& \textbf{Para.} & \textbf{Reduced Model} & \textbf{Full Model} \\
	\hline
	& & Median (Range) & Median (Range) \\
	\hline
	\hline
	\multirow{4}{*}{\textbf{Mean Vector}} 
	& $\mu_{\eta_{0}}$ & $0.0026$ ($0.0019$, $0.0035$) & $0.0026$ ($0.0019$, $0.0035$) \\
	& $\mu_{\eta_{1}}$ & $-0.0128$ ($-0.0178$, $-0.0078$) & $-0.0128$ ($-0.0178$, $-0.0078$) \\
	& $\mu_{\eta_{2}}$ & $-0.0118$ ($-0.0498$, $-0.0050$) & $-0.0120$ ($-0.0490$, $-0.0050$) \\
	& $\mu_{\gamma}$ & $0.0092$ ($0.0036$, $0.0190$) & $0.0092$ ($0.0037$, $0.0204$) \\
	\hline
	\hline
	\multirow{4}{*}{\textbf{Path Coef. of $x_{1}$}} 
	& $\beta_{10}$ & $0.2594$ ($0.1595$, $0.4266$) & $0.2568$ ($0.1570$, $0.4227$) \\
	& $\beta_{11}$ & $0.2980$ ($0.1656$, $0.5418$) & $0.2752$ ($0.1678$, $0.4674$) \\
	& $\beta_{12}$ & $0.2912$ ($0.1634$, $0.5378$) & $0.2734$ ($0.1646$, $0.4641$) \\
	& $\beta_{1\gamma}$ & ---\tnote{1} & $0.7689$ ($0.3162$, NA\tnote{2}) \\
	\hline
	\hline
	\multirow{4}{*}{\textbf{Path Coef. of $x_{2}$}} 
	& $\beta_{20}$ & $0.1747$ ($0.1063$, $0.2865$) & $0.1722$ ($0.1059$, $0.2830$) \\
	& $\beta_{21}$ & $0.2234$ ($0.1104$, $0.4097$) & $0.1860$ ($0.1114$, $0.3068$) \\
	& $\beta_{22}$ & $0.2234$ ($0.1091$, $0.4122$) & $0.1834$ ($0.1104$, $0.3061$) \\
	& $\beta_{2\gamma}$ & --- & $0.5625$ ($0.2179$, NA) \\
	\hline
	\hline
	\multirow{4}{*}{\textbf{Unexplained Variance}} 
	& $\psi_{00}$ & $0.0856$ ($0.0610$, $0.1111$) & $0.0846$ ($0.0610$, $0.1112$) \\
	& $\psi_{11}$ & $0.1076$ ($0.0655$, $0.1998$) & $0.0982$ ($0.0664$, $0.1360$) \\
	& $\psi_{22}$ & $0.1078$ ($0.0663$, $0.2038$) & $0.0988$ ($0.0666$, $0.1326$) \\
	& $\psi_{\gamma\gamma}$ & --- & $0.6089$ ($0.2130$, NA) \\
	\hline
	\hline
	\end{tabular}
	\label{tbl:rRMSE}
	\begin{tablenotes}
	\small
	\item[1] {--- indicates that the relative RMSEs are not available from the reduced model.}
	\item[2] {NA indicates that the upper bound of relative RMSE is not available. The model performance under the conditions with $0$ population value of difference in knot is of interest where the relative RMSE of those knot coefficients would go infinity.} \\
	\end{tablenotes}
	\end{threeparttable}
\end{table}

\begin{table}[ht]
	\centering
	\begin{threeparttable}
	\setlength{\tabcolsep}{5pt}
	\renewcommand{\arraystretch}{0.8}
	\caption{Median (Range) of Coverage Probability of Each Parameter Across Conditions with $10$ Repeated Measures \& Midway Knot}
	\begin{tabular}{llrr}
	\hline
	\hline
	& \textbf{Para.} & \textbf{Reduced Model} & \textbf{Full Model} \\
	\hline
	& & Median (Range) & Median (Range) \\
	\hline
	\hline
	\multirow{4}{*}{\textbf{Mean Vector}} 
	& $\mu_{\eta_{0}}$ & $0.9470$ ($0.9280$, $0.9680$) & $0.9459$ ($0.9203$, $0.9677$)\tnote{1} \\
	& $\mu_{\eta_{1}}$ & $0.9430$ ($0.7800$, $0.9660$) & $0.9418$ ($0.7660$, $0.9744$) \\
	& $\mu_{\eta_{2}}$ & $0.9440$ ($0.7970$, $0.9670$) & $0.9429$ ($0.7990$, $0.9702$) \\
	& $\mu_{\gamma}$ & $0.9100$ ($0.7880$, $0.9680$) & $0.9510$ ($0.9267$, $0.9758$) \\
	\hline
	\hline
	\multirow{4}{*}{\textbf{Path Coef. of $x_{1}$}} 
	& $\beta_{10}$ & $0.9460$ ($0.9140$, $0.9620$) & $0.9472$ ($0.8926$, $0.9745$) \\
	& $\beta_{11}$ & $0.9295$ ($0.6170$, $0.9630$) & $0.9450$ ($0.9070$, $0.9673$) \\
	& $\beta_{12}$ & $0.9290$ ($0.6120$, $0.9630$) & $0.9460$ ($0.9080$, $0.9823$) \\
	& $\beta_{1\gamma}$ & ---\tnote{2} & $0.9463$ ($0.6640$, $0.9751$) \\
	\hline
	\hline
	\multirow{4}{*}{\textbf{Path Coef. of $x_{2}$}} 
	& $\beta_{20}$ & $0.9440$ ($0.8950$, $0.9590$) & $0.9479$ ($0.9220$, $0.9685$) \\
	& $\beta_{21}$ & $0.9090$ ($0.2780$, $0.964$) & $0.9440$ ($0.8820$, $0.9745$) \\
	& $\beta_{22}$ & $0.9130$ ($0.2960$, $0.9650$) & $0.9445$ ($0.8720$, $0.9703$) \\
	& $\beta_{2\gamma}$ & --- & $0.9416$ ($0.3810$, $0.9879$) \\
	\hline
	\hline
	\multirow{4}{*}{\textbf{Unexplained Variance}} 
	& $\psi_{00}$ & $0.9375$ ($0.8970$, $0.9600$) & $0.9389$ ($0.9086$, $0.9627$) \\
	& $\psi_{11}$ & $0.9365$ ($0.4240$, $0.9600$) & $0.9351$ ($0.8550$, $0.9728$) \\
	& $\psi_{22}$ & $0.9370$ ($0.4660$, $0.9620$) & $0.9360$ ($0.8570$, $0.9698$) \\
	& $\psi_{\gamma\gamma}$ & --- & $0.9620$ ($0.0030$, $0.9976$) \\
	\hline
	\hline
	\end{tabular}
	\label{tbl:CP}
	\begin{tablenotes}
	\small
	\item[1] {For the full model, the reported coverage probabilities have four decimals since we calculated the coverage probabilities only based on the replications with proper solutions.}
	\item[2] {--- indicates that the coverage probabilities are not available from the reduced model.}
	\end{tablenotes}
	\end{threeparttable}
\end{table}

\begin{table}[ht]
	\centering
	\begin{threeparttable}
	\setlength{\tabcolsep}{5pt}
	\renewcommand{\arraystretch}{0.8}
	\caption{Summary of Model Fit Information For the Models}
	\begin{tabular}{lrrrrr}
	\hline
	\hline
	\multicolumn{6}{c}{\textbf{Latent Growth Curve Models (Time Structure: Age in Months)}} \\
	\hline
	\textbf{Model} & \textbf{-2loglikelihood} & \textbf{AIC}  & \textbf{BIC}  & \textbf{\# of Para.} & \textbf{Residual}  \\
	\hline
	Linear & $27298$ & $27310$ & $27334$ & $6$ & $75.60$ \\
	\hline
	Quadratic & $25320$ & $25340$ & $25379$ & $10$ & $35.14$ \\
	\hline
	Jenss-Bayley & $25330$ & $25352$ & $25396$ & $11$ & $35.37$ \\
	\hline
	Linear-linear Piecewise w/ a Fixed Knot & $25412$ & $25434$ & $25477$ & $11$ & $36.84$ \\
	\hline
	Linear-linear Piecewise w/ a Random Knot & $25308$ & $25338$ & $25398$ & $15$ & $33.06$ \\
	\hline
	\hline
	\multicolumn{6}{c}{\textbf{Latent Growth Curve Models with TICs (Time Structure: Age in Months)}} \\
	\hline
	\textbf{Model} & \textbf{-2loglikelihood} & \textbf{AIC}  & \textbf{BIC}  & \textbf{\# of Para.} & \textbf{Residual}  \\
	\hline
	Proposed Model & $31582$ & $31714$ & $31977$ & $66$ & $33.00$ \\
	\hline
	\hline
	\multicolumn{6}{c}{\textbf{Latent Growth Curve Models with TICs (Time Structure: Grade-in-school in Months)}} \\
	\hline
	\textbf{Model} & \textbf{-2loglikelihood} & \textbf{AIC}  & \textbf{BIC}  & \textbf{\# of Para.} & \textbf{Residual}  \\
	\hline
	Proposed Model & $31580$ & $31912$ & $31975$ & $66$ & $34.28$ \\
	\hline
	\hline
	\end{tabular}
	\label{tbl:info}
	\end{threeparttable}
\end{table}

\begin{table}[ht]
	\centering
	\begin{threeparttable}
	\setlength{\tabcolsep}{5pt}
	\renewcommand{\arraystretch}{0.8}
	\caption{Estimates of the Parameters of Mathematics Trajectories}
	\begin{tabular}{p{6.6cm}rrr|rrr}
	\hline
	\hline
	\textbf{Growth Factor} & \multicolumn{3}{c}{\textbf{Intercept}} & \multicolumn{3}{c}{\textbf{First Slope}} \\
	\hline
	\textbf{Parameter} & \textbf{Estimate} & \textbf{SE} & \textbf{P value} & \textbf{Estimate} & \textbf{SE} & \textbf{P value} \\
	\hline
	\textbf{Mean} & $26.855$ & $0.589$ & $<0.0001^{\ast}$ &	$1.731$ & $0.019$ &	$<0.0001^{\ast}$ \\
	\textbf{Unexplained Variance} & $106.321$ &	$9.822$ & $<0.0001^{\ast}$ & $0.083$ & $0.010$ & $<0.0001^{\ast}$ \\
	\textbf{Sex}($0-$Boy; $1-$Girl) & $0.377$ &	$0.591$	& $0.5235$ & $-0.042$ & $0.019$ & $0.0271^{\ast}$ \\
	\textbf{Race}($0-$White; $1-$Other) & $-0.245$ & $0.685$ & $0.7206$	& $-0.001$ & $0.022$ & $0.9637$ \\
	\textbf{School Location} & $-0.585$	& $0.634$ &	$0.3562$ & $-0.027$ & $0.020$ & $0.1770$ \\
	\textbf{Income} & $1.743$ &	$0.838$ & $0.0375^{\ast}$ & $0.053$ & $0.027$ &	$0.0497^{\ast}$ \\
	\textbf{School Type}($0-$Public; $1-$Private) & $1.841$ & $0.611$ & $0.0026^{\ast}$	& $-0.025$ & $0.019$ &	$0.1882$ \\
	\textbf{Parents' Highest Education} & $2.863$ &	$0.787$	& $0.0003^{\ast}$ & $0.045$ & $0.025$ &	$0.0719$	 \\
	\hline
	\hline
	\textbf{Growth Factor} & \multicolumn{3}{c}{\textbf{Second Slope}} & \multicolumn{3}{c}{\textbf{Knot}} \\
	\hline
	\textbf{Parameter}& \textbf{Estimate} & \textbf{SE} & \textbf{P value} & \textbf{Estimate} & \textbf{SE} & \textbf{P value} \\
	\hline
	\textbf{Mean} & $0.688$ & $0.017$ & $<0.0001^{\ast}$ & $101.741$ & $0.523$ & $<0.0001^{\ast}$ \\
	\textbf{Unexplained Variance} & $0.014$ & $0.008$ &	$0.0801$ & $42.477$ & $8.308$ & $<0.0001^{\ast}$ \\
	\textbf{Sex}($0-$Boy; $1-$Girl) & $0.021$ &	$0.017$ & $0.2167$ & $-0.805$ & $0.535$ & $0.1324$ \\
	\textbf{Race}($0-$White; $1-$Other) & $-0.043$ & $0.019$ & $0.0236^{\ast}$ & $0.196$ &	$0.603$	& $0.7451$ \\
	\textbf{School Location} & $-0.023$ & $0.018$ & $0.2013$ & $1.692$ & $0.559$ & $0.0025^{\ast}$ \\
	\textbf{Income} & $0.008$ &	$0.025$ & $0.7490$ & $-0.529$ & $0.799$ & $0.5079$ \\
	\textbf{School Type}($0-$Public; $1-$Private) & $-0.028$ & $0.018$ & $0.1198$ & $0.097$ & $0.554$ & $0.8610$ \\
	\textbf{Parents' Highest Education} & $-0.026$ & $0.023$ & $0.2583$ & $0.337$ &	$0.739$ & $0.6484$ \\
	\hline
	\hline
	\end{tabular}
	\label{tbl:est}
	\begin{tablenotes}
	\small
	\item[1] $^{\ast}$ indicates statistical significance at $0.05$ level.
	\end{tablenotes}
	\end{threeparttable}
\end{table}

\newpage
%

\renewcommand\thefigure{A.\arabic{figure}}
\setcounter{figure}{0}
\begin{figure}[ht]
	\centering
	\includegraphics[width=1.0\textwidth]{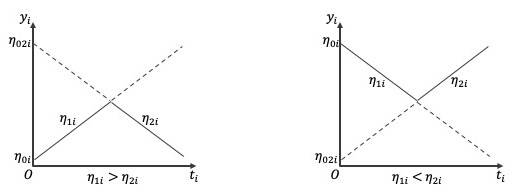}
	\caption{The Two Forms of the Bilinear Spline (Linear-Linear Piecewise)}
	\label{fig:proj1_2cases}
\end{figure}

\end{document}